\documentclass[12pt]{iopart}
\usepackage{amssymb}
\usepackage{graphicx}
\usepackage{bm}
\begin{document}
\title{Active and passive fields face to face}

\author{Antonio Celani\dag\footnote[2]{To whom correspondence should
be addressed (Antonio.Celani@inln.cnrs.fr)}, Massimo Cencini\P,
Andrea Mazzino\S and Massimo Vergassola$\|$}

\address{\dag\ CNRS, INLN, 1361 Route des Lucioles, 06560 Valbonne,
France.}  
\address{\P\ Center for Statistical Mechanics and Complexity,
INFM Roma 1 and Dipartimento di Fisica, Universit\`a di
Roma "La Sapienza", Piazzale Aldo Moro, 2, I-00185 Roma, Italy}
\address{\S\ INFM--Dipartimento di Fisica, Universit\`a di Genova, Via
Dodecaneso 33, I-16146 Genova, Italy. } 
\address{$\|$ CNRS,
Observatoire de la C\^ote d'Azur, B.P. 4229, 06304 Nice Cedex 4,
France, and CNRS, URA 2171, Institut Pasteur, 28 rue du Dr. Roux,
75724 Paris Cedex 15, France.}
\begin{abstract}
The statistical properties of active and passive scalar fields
transported by the same turbulent flow are investigated.  Four
examples of active scalar have been considered: temperature in thermal
convection, magnetic potential in two-dimensional
magnetohydrodynamics, vorticity in two-dimensional Ekman turbulence and
potential temperature in surface flows. In the cases of temperature and
vorticity, it is found that the active scalar behavior is akin to that of its
co-evolving passive counterpart.  The two other cases indicate that this similarity 
is in fact not generic and differences between passive and active fields can 
be striking: in two-dimensional
magnetohydrodynamics the magnetic potential performs an inverse
cascade while the passive scalar cascades toward the small-scales; in
surface flows, albeit both perform a direct cascade, the potential
temperature and the passive scalar have different scaling laws already
at the level of low-order statistical objects.  These dramatic
differences are rooted in the correlations between the active scalar
input and the particle trajectories. The role of such correlations in
the issue of universality in active scalar transport and the behavior
of dissipative anomalies is addressed.
\end{abstract}
\pacs{47.27.-i}
\submitto{\NJP}
\maketitle 

\section{Introduction}
\label{sec:intro}
Scalar fields transported by turbulent flows are encountered in many
natural phenomena and engineering problems, ranging from atmospheric
physics \cite{PS83} to combustion \cite{W85}, to the transport and
amplification of magnetic fields in astrophysical fluids \cite{ZRS83}.
Here, we consider the case of advected scalar fields such as
temperature, pollutant density, chemical or biological species
concentration.

In many cases there is  a two-way coupling
between the scalar and the flow: the transported field can 
influence the velocity field -- this is dubbed {\it active} transport.
This is the case, for example, of the temperature field that acts on
velocity through buoyancy forces. Conversely, situations where the 
feedback of the scalar field is negligible and
the velocity determines the properties of the scalar, but not {\it vice versa} 
are known as {\it passive}. This ideal case is well approximated by fluorescent
dye used in laboratory experiments to mark fluid parcels.

Although active and passive scalars are governed by the same 
advection-diffusion equation, their nature is radically different.  Passive
scalars belong to the realm of linear problems, even though highly
nontrivial.  Indeed, as a consequence of the statistical independence
of the forcing and the advecting velocity, the transported field depends
linearly on the forcing.  This property allows a full-fledged
theoretical treatment of the problem, and has the major consequence
that the passive scalar scaling laws are universal with respect to
the injection mechanism.  On the contrary, for active fields the
presence of the feedback couples the velocity to the transported
scalar and makes the problem fully nonlinear. In this case, the
theoretical tools developed for the study of the passive problem may
fall short of explaining the behavior of active scalars, and, as of
today, the understanding of active turbulent transport lags far behind
the knowledge accumulated for the passive counterpart.  This state of
the art motivated us to pursue a ``case study'' of turbulent transport
of active and passive scalars, using the scaling properties of fields
evolving in the same turbulent flow as the basic diagnostics for
comparison.

We consider four different systems belonging to the
two following general classes of problems: {\it (i)\/} 
active scalars that influence
the flow through local forces; {\it (ii)} 
active fields functionally related to
the velocity.

The evolution of a scalar belonging to the 
first class is described in terms of the following set of
equations:
\begin{eqnarray}
\partial_t a+ {\bm v}\cdot {\bm \nabla} a& = &\kappa \Delta\, a + f_a \;,
\label{eq:1}\\
\partial_t c+ {\bm v}\cdot {\bm \nabla} c& = &\kappa \Delta\, c + f_c \;,
\label{eq:2}\\
\partial_t {\bm v} + {\bm v}\cdot {\bm \nabla} {\bm v}&=& -{\bm \nabla} p
+ \nu\Delta{\bm v} + {\bm F}[a,{\bm \nabla} a,\dots]\;,
\label{eq:3}
\end{eqnarray}
where $a$ and $c$ are the active and passive scalar fields,
respectively and obey the advection-diffusion equations (\ref{eq:1})
and (\ref{eq:2}). The scalar inputs $f_a$ and $f_c$ have a
characteristic lengthscale $\ell_f$, and represent two different
realizations of the same stochastic process.  Were they coincident, no
difference between active and passive fields would persist.  For the
sake of simplicity, we use the same molecular diffusivity, $\kappa$
for both scalars.  The active character of $a$ is embodied by the term
${\bm F}[a,{\bm \nabla} a,\dots]$, which acts as a forcing for the
velocity field in the Navier-Stokes equations.  The specific form of
the ${\bm F}$ term depends on the physical system under
investigation. In the following we shall consider two examples from
this class: thermal convection \cite{MY75,S94}, where $a$ is the
temperature field and ${\bm F}=-\beta a {\bm g}$ is the buoyancy
force, and two-dimensional magnetohydrodynamics \cite{B93}, where $a$
is the magnetic potential and ${\bm F} = -\Delta a{\bm \nabla}a$ is
the Lorentz force.

The second class of active scalars is relevant to
geophysical flows \cite{S98,R79}. In this case the dynamics
is described in terms of a scalar field obeying the
advection-diffusion equation (\ref{eq:1}), supplemented by 
a functional relation that gives ${\bm v}$ in terms of $a$ :
\begin{equation}
v_i({\bm x},t)=\int \Gamma_i({\bm x}-{\bm y}) a({\bm y},t) {\rm d}{\bm y}\,.
\label{eq:4}
\end{equation}
Here, the vector valued kernel ${\bm \Gamma}$ is divergence-free.  A
well-known instance from such class is the two-dimensional
Navier-Stokes equation, where the active scalar is the vorticity, ${\bm
\nabla} \times {\bm v}$. Another problem which we shall discuss is the
turbulent flow on the flat surface of an infinitely high fluid, described 
by the surface-quasi-geostrophic equation
\cite{PHS94,C98}.  Here, the active scalar is the fluid density, which
is related (e.g. for ideal gases) to the so-called potential
temperature.

Some of the results presented in this paper were previously discussed in
Refs.~\cite{CMMV02,BCMV02,CCMV02}. Related investigations on 
active and passive transport may be found in 
\cite{CCGP02,CCGP03} and \cite{NAGO99,NAGO00}.

The paper is organized as follows. 
In Sect.~\ref{sec:1}, we briefly review some results about passive
scalar transport in turbulent flows. 
Particular emphasis is put on the Lagrangian description
of scalar transport, pointing out the results which hold
for active scalars as well.  In Sect.~\ref{sec:2} the statistics of the
(active) temperature and a passive scalar field in $2d$
convection is discussed.  Sect.~\ref{sec:3} is devoted to the analysis
of two-dimensional magnetohydrodynamics, with a thorough 
discussion of the problem of dissipative anomaly in scalar transport. In
Sect.~\ref{sec:4} the evolution of passive and active fields in 
Ekman-Navier-Stokes turbulence is discussed.  
In Sect.~\ref{sec:5} we study the properties of
turbulence on fluid surfaces under the quasi-geostrophic approximation.
In the last section we summarize the main findings and 
delineate the perspectives for active scalar transport. 

\section{Passive scalar turbulence}
\label{sec:1}
\subsection{Eulerian approach}

The dynamics of passive scalars is governed by the advection diffusion equation
(\ref{eq:2}). In order to describe the general properties of passive
scalar evolution in turbulent incompressible velocity fields,
we assume that the velocity field ${\bm v}$ is scale invariant and rough.
In other words, the spatial increments $\delta_r v\!=\!({\bm v}({\bm
x}\!+\!{\bm r},t)\!-\!{\bm v}({\bm x},t))\cdot {\bm r}/r$ depend on the
separation $r$ as a
fractional power, i.e. $\delta_r v\sim r^{h}$ with $h<1$
(e.g. $h=1/3$ in Kolmogorov's 1941 turbulence \cite{F95}).  Being 
interested in the statistically steady properties of the field, we
introduce a source of scalar fluctuations $f_c$. 
In the following we take for convenience a random, Gaussian, statistically 
homogeneous and isotropic forcing with zero mean and correlation function
\begin{equation}
\langle f_c({\bm x}_1,t)f_c({\bm
x}_2,t')\rangle=\delta(t-t'){\cal F}(|{\bm x}_1-{\bm x}_2|/\ell_f)\,.
\label{eq:corrf}
\end{equation}
The correlation function of the forcing 
 ${\cal F}(r/\ell_f)$ is roughly constant at scales smaller than
$\ell_f$, which is assumed to be within the scaling range of ${\bm v}$, 
and drops
rapidly to zero for $r>\ell_f$. 

The phenomenology of passive scalar turbulence may be summarized as
follows. Scalar fluctuations injected at the scale $\ell_f$ are 
transferred toward the small scales
with a constant flux, down to the dissipative scale $\ell_d$.
There, the molecular diffusion absorbs the incoming flux and ensures
the equilibrium between the input and the dissipation. The fluctuations
are thus maintained in a statistically steady state which is
characterized by two major properties. First, the scalar
dissipation is asymptotically independent of the
molecular diffusivity $\kappa$, attaining a finite nonzero limiting
value for $\kappa \to 0$. This singular behavior of the dissipation
is also known as {\it dissipative anomaly}.  Second, in the scaling
range $\ell_d \ll r \ll \ell_f$ the scalar statistics is {\it
intermittent}.  This amounts to saying that the small-scale statistics
is characterized by the alternance of strong, rare events where the
scalar increments, $\delta_r c\!=\! c({\bm x}\!+\!{\bm
r},t)\!-\!c({\bm x},t)$ are much larger than their typical value,
$c_{rms}\,$, and long quiescent phases where $\delta_r c \ll
c_{rms}$. Intermittency is reflected by the
scaling behavior of the structure functions, i.e. the moments of the
scalar increments
\begin{equation}
S^c_N(r)=\langle (\delta_r c)^N\rangle \propto r^{\zeta^{dim}_N}
\left( {\ell_f \over r}\right)^{\zeta^{dim}_N-\zeta^c_N}\,.
\label{eq:sf}
\end{equation}
The scaling exponents $\zeta^c_N$ are said to be {\it anomalous}, when
they deviate from the dimensional expectation
$\zeta^{dim}_N=N(1-h)/2$.  The equality $\zeta^c_N=\zeta^{dim}_N$
holds possibly only for $N=2$ \cite{MY75}, whereas for $N>2$ the
deviations become more and more severe.  The asymptotic behavior at 
large orders $N$ corresponds to the saturation 
$\zeta^c_N\to \zeta^c_{\infty}$
\cite{CLMV01}. The saturation is related to the presence of sharp
``fronts'' in the scalar field. The exponents $\zeta^c_N$ are {\it universal}
with respect to the details of the energy injection statistics. 
The forcing only affects the numerical prefactors appearing in
the structure functions.

The anomalous scaling $\zeta^{dim}_N\neq\!  \zeta^c_N$ signals the
breakdown of scale invariance, as confirmed by the explicit appearance
of $\ell_f$ in (\ref{eq:sf}), even at scales $r\!\ll\!  \ell_f$. Indeed,
anomalous scaling of the moments of scalar increments is equivalent to
state that the probability density functions (pdf's) of $\delta_r c$ at
different $r$'s cannot be collapsed by rescaling them with a unique
power law in $r$. Even though the specific values of the exponents
$\zeta^c_N$ depend on the details of the flow (statistics, time correlation,
and roughness exponent $h$), intermittency and the breaking of scale
invariance in the scalar statistics are generic features of passive
scalar turbulence.

The physical mechanism leading to anomalous scaling has been recently
understood in the framework of the Kraichnan
model of scalar advection \cite{K68,K94} 
(see Refs.~\cite{GK95,CFKL95,SS95} and  Ref.~\cite{FGV01} 
for an exhaustive review on the subject). In this model 
the advecting flow ${\bm v}$ is random,
Gaussian, self-similar and $\delta$-correlated in time. 
Under these special conditions,
 there exists a closed set of linear equations for multi-point correlation
functions. 
The anomalous exponents are the scaling exponents of the homogeneous
solutions (the so-called {\it zero modes}) of those equations. Since 
homogeneous solutions do not depend by definition on the scalar input, 
their scaling
exponents are universal, and cannot be inferred from dimensional arguments. 
The concept of zero mode can be extended to passive scalar turbulence in
generic velocity fields \cite{CV01}. 

The properties of passive scalars described up to now are 
in the language of
fields --the Eulerian description. It is now
interesting to adopt a different but equivalent viewpoint in terms of
particle trajectories --the Lagrangian description. 
\subsection{Lagrangian description}
\label{sec:lag}
The basic idea of the Lagrangian approach is to solve Eq.~(\ref{eq:2})
by the method of characteristics. Let us denote ${\bm \rho}(s;{\bm
x},t)$ (henceforth whenever there is no ambiguity we indicate it
as ${\bm \rho}(s)$) as the trajectory of a fluid particle 
landing at point ${\bm x}$ at time $t$.
The path ${\bm \rho}(s)$ is the
solution of the stochastic differential equation
\begin{equation}
{d{\bm \rho}(s) \over ds}= {\bm v}({\bm
\rho}(s),s)+\sqrt{2\kappa}\,\dot{\bm w}(s)\,,\qquad {\bm \rho}(t) = {\bm x}\,,
\label{eq:traj}
\end{equation}
where $i,j=1,\dots,d$ ($d$ being the space dimensionality), and
$\dot{\bm w}(s)$ is a Wiener process (the derivative of a Brownian
motion), i.e.  $\dot{w}_i$ are Gaussian variables of zero mean and
correlation $\langle \dot{w}_i(s)\dot{w}_j(s')
\rangle=\delta_{ij}\delta(s-s')$.  Along the path ${\bm \rho}(s)$
Eq.~(\ref{eq:2}) reduces to
\begin{equation}
{d\phi^{\bm w}(s)\over ds}= f_c({\bm \rho}(s),s)\,,
\label{eq:caracteristics}
\end{equation}
which is easily solved as $\phi^{\bm w}(t)=\int_0^t {\rm d}s\,
f_c({\bm \rho}(s),s)$. For the sake of simplicity, we assumed
$\phi^{\bm w}(0)=0$.  We indicated with $\phi^{\bm w}$ the solution
obtained along the path ${\bm \rho}$ obtained for a specific
realization of the process ${\bm w}$. The passive scalar
field, $c({\bm x},t)$, is recovered by averaging over all the
realizations of ${\bm w}$, i.e. along all the Lagrangian 
paths ending in ${\bm x}$ at time $t$ \cite{FMNV99}:
\begin{equation}
c({\bm x},t)=\langle \phi^{\bm w}(t) \rangle_{\bm w}=\left\langle \int_0^t
{\rm d}s\, f_c({\bm \rho}(s),s) \right\rangle_{\bm w}\,.
\label{eq:lag}
\end{equation}

The statistic of the trajectories is summarized in the particle 
propagator $P({\bm
y},s|{\bm x},t)=\langle \delta({\bm y}-{\bm \rho}(s;{\bm
x},t)\rangle_{\bm w}$, that is the probability of finding a
particle at point ${\bm y}$ and time $s \le t$, provided it is in
${\bm x}$ at time $t$.  According to the theory of stochastic processes
\cite{Risk}, $P({\bm y},s|{\bm x},t)$ obeys the Kolmogorov equations:
\begin{eqnarray}
- \partial_s P({\bm y},s|{\bm x},t) -{\bm \nabla}_{\!y} \cdot 
[{\bm v}({\bm y},s) P({\bm y},s|{\bm x},t) ] \!&=&\! 
\kappa \Delta_y P({\bm y},s|{\bm x},t) \,,
\label{eq:Kolmo1}\\
\;\;\;\partial_t P({\bm y},s|{\bm x},t) +{\bm \nabla}_{\!x} \cdot 
[{\bm v}({\bm x},t) P({\bm y},s|{\bm x},t) ] \!&=&\! 
\kappa \Delta_x P({\bm y},s|{\bm x},t) \,,
\label{eq:Kolmo2}
\end{eqnarray} 
with initial condition 
$P({\bm y},t|{\bm x},t)=\delta({\bm x}-{\bm y})$. The unusual minus
signs in the l.h.s. of (\ref{eq:Kolmo1}) are due to the fact that
particles move {\em backward} in time.  The solution of (\ref{eq:2})
can be written in terms of the propagator:
\begin{equation}
c({\bm x},t)=\int_0^t ds\int d{\bm y}\,f_c({\bm y},s) 
P({\bm y},s|{\bm x},t)\;,
\label{eq:plag}
\end{equation} 
as it can be directly checked by inserting (\ref{eq:plag}) in (\ref{eq:2}) 
and using (\ref{eq:Kolmo2}). 

At variance with smooth velocities (i.e. Lipschitz continuous,
$\delta_r v\sim r$) where for $\kappa \to 0$ particle trajectories are
unique, for velocity fields rough and incompressible the particle
propagator does not collapse onto a single trajectory in the limit
$\kappa \to 0$.  Lagrangian paths are not unique and initially
coincident particles separate in a finite time.  This property is at the
root of the dissipative anomaly. 

For active and passive scalars evolving in the same flow, the
Lagrangian paths ${\bm \rho}(s)$ are the same, as well as the
propagator $P({\bm y},s|{\bm x},t)$.  The difference between $a$ and
$c$ is that, since the active scalar enters the dynamics of ${\bm v}$
(\ref{eq:3}-\ref{eq:4}), the Lagrangian trajectories are functionally
related to the active scalar forcing $f_a$, but are independent of the
passive source term $f_c$. This is the Lagrangian counterpart of the
linearity of the passive scalar problem, which does not hold for the
more complicated case of active scalars.

One of the advantages of the Lagrangian description is that it makes
transparent the physics of transport processes.  For instance, let us
consider the $2$-points correlation function for passive scalars,
${\cal C}^{c}({\bm x}_1,{\bm x}_2;t)=\langle c({\bm x}_1,t)c({\bm
x}_2,t) \rangle$.  This offers the possibility of an intuitive
interpretation of the energy cascade phenomenology and gives insights
into measurable statistical objects such as the scalar spectrum
$E_c(k)$, that is the Fourier transform of ${\cal C}^c_2$. From
(\ref{eq:lag}) and averaging over $f_c$ and ${\bm v}$
one obtains
\begin{equation}
{\cal C}^c_2({\bm x}_1,{\bm x_2};t)= \left \langle \int_0^t \!{\rm d}s_1 \int_0^t \!{\rm d}s_2 \left\langle f_c({\bm \rho}(s_1;{\bm x}_1,t))f_c({\bm \rho}(s_2;{\bm x}_2,t))
\right\rangle_f \right\rangle_{{\bm w}{\bm v}}\,.
\label{eq:c2lag} 
\end{equation}
Introducing the velocity-averaged two-particle propagator $\langle
P_2({\bm y}_1,{\bm y}_2,s|{\bm x}_1,{\bm x}_2,t)\rangle_v$, which
evolves according to the straightforward generalization of
(\ref{eq:Kolmo1}, \ref{eq:Kolmo2}) to two particles,
and using (\ref{eq:corrf}),  Eq.~(\ref{eq:c2lag}) reduces to
\begin{equation}
\fl \qquad {\cal C}^c_2({\bm x}_1,{\bm x_2};t)=\int_0^t {\rm d}s \int \int
\langle P_2({\bm y}_1,{\bm y}_2,s|{\bm x}_1,{\bm x}_2,t)\rangle_v 
{\cal F}\left({|{\bm y}_1-{\bm y}_2|/ \ell_f}\right) {\rm d}{\bm y}_1
{\rm d}{\bm y}_2 \,,
\label{eq:c2plag}
\end{equation}
which has a clear physical interpretation. Since ${\cal F}(x)$ is
vanishingly small for $x\gg 1$, the correlation ${\cal C}^c_2$ is essentially the
average time spent by a particle pair with a separation $r=|{\bm
x}_1-{\bm x}_2|$ below the forcing scale $\ell_f$.  Due to the
explosive separation of particles, this time has a finite limit for $r
\to 0$, which yields the leading contribution to ${\cal C}^c_2$.
The subleading behavior is uncovered by the second-order
structure function
\begin{equation}
S^c_2(r)=\langle (c({\bm
x}_1,t)-c({\bm x}_2,t))^2\rangle=2({\cal C}^c_2(0)-{\cal C}^c_2(r))\,,
\label{eq:s2}
\end{equation}
which is roughly the average time ${\cal T}_{\ell_f}(r)$ taken by
two coinciding particles to reach a separation $r$.  For a Kolmogorov
1941 turbulent flow ($h=1/3$) one has
$S^c_2(r)\sim r^{2/3}$, i.e. $E_c(k)\sim k^{-5/3}$  -- the
Oboukov-Corrsin dimensional expectation \cite{MY75}.

The Lagrangian description can be extended also to higher-order statistics as
multipoint correlation functions ${\cal C}^c_N({\bm x}_1,\ldots,{\bm
x}_N)= \langle c({\bm x}_1,t) \ldots c({\bm x}_N,t) \rangle$.
However, when many points come into play, their geometrical arrangement  becomes
crucial. Dimensional arguments, which are based on the size 
of the configuration but forcibly neglect the ``angular'' information,
fall short of capturing the 
observed behavior for multi-point observables.
A detailed discussion of their properties 
is beyond the scope of this brief review. In the following we 
just summarize the main concepts, referring
to Refs.~\cite{BGK98,FGV01} for further reading. 

Expanding the power in the definition of the structure functions 
$S^c_N(r)$, it is immediate to express them as a linear 
combinations of $N$-point correlation functions (see, e.g., (\ref{eq:s2}) for
$S^c_2(r)$). Therefore, the latter must contain a contribution,
denoted as ${\cal Z}^c_N({\bm x}_1,\ldots,{\bm x}_N)$, that carries
the anomalous scale dependence: ${\cal
Z}^c_N(\lambda{\bm x}_1,\ldots,\lambda{\bm x}_N)=\lambda^{\zeta^c_N}
{\cal Z}^c_N({\bm x}_1,\ldots,{\bm x}_N)$ \cite{GK95,CFKL95,SS95}.
From a Lagrangian viewpoint, the function ${\cal Z}^c_N$ has a special
property that distinguishes it from a generic scaling
function. The remarkable result is that \cite{BGK98,CV01}
\begin{equation}
\frac{d}{dt} \langle {\cal Z}^c_N \rangle_{\cal L}= 0\,,
\label{eq:sp}
\end{equation}
where the derivative $\frac{d}{dt}$ is performed along the trajectories
of $N$ particles advected by the flow, and the average is over the ensemble
of all trajectories. In other terms, ${\cal Z}^c_N$ is {\em
statistically preserved} by the flow \cite{CV01,ABCPV01}. 
Universality of scaling exponents is then just
a byproduct of the definition of statistically preserved
structures: since $f_c$ does not appear in Eq.~(\ref{eq:sp}), 
the properties of zero modes are insensitive to the choice of the forcing.

\subsection{Dissipative anomaly}
\label{sec:diss}

In spite of the continuous injection of scalar through the pumping $f_c$, 
the second-order moment $\langle c^2({\bm x},t) \rangle$ 
does not grow indefinitely even in the limit $\kappa \to 0$.
This is due to the existence of a finite nonzero limit of 
the scalar dissipation 
$\epsilon_c = \kappa |{\bm \nabla c}|^2$ -- the dissipative anomaly.

In order to understand how $\langle c^2({\bm x},t) \rangle$ 
achieves a finite value independent of the diffusivity coefficient,
we adopt the Lagrangian viewpoint. From Eq.~(\ref{eq:c2lag})
 we have:
\begin{equation}
\fl \langle c^2({\bm x},t)\rangle\!\! =\!\! \left\langle
\int_0^t\int_0^t{\rm d}s_1 {\rm d}s_2 f_c({\bm \rho}(s_1;{\bm x},t))
f_c({\bm \rho}(s_2;{\bm x},t)) \right \rangle
=\left\langle \left(\int_0^t {\rm d}s f_c({\bm \rho}(s;{\bm
x},t))\right)^2\right\rangle \,,\label{eq:csquaredlag}
\end{equation}
where the brackets indicate the average over the scalar forcing, the
velocity field and the noise.

Looking na\"{\i}vely at (\ref{eq:csquaredlag}) one might expect that for
a large class of random forcing of zero mean the r.h.s of the above
expression would grow linearly with $t$.  For instance, when the
forcing is Gaussian and $\delta$-correlated in time, one could argue
that (\ref{eq:csquaredlag}) is essentially a sum of independent
variables and by central limit theorem arguments conclude that 
$\langle c^2\rangle \propto t$.  This conclusion would be correct if
in the limit $\kappa \to 0$ all trajectories collapse onto a unique
Lagrangian path.  This turns out to be the case for strongly compressible
flows but not in general. For compressible flows, energy indeed 
grows linearly in time and the
advected scalar performs an inverse cascade process
\cite{CKV97,GV00}. On the contrary, in rough incompressible flows
coinciding particles typically separate in a finite time, giving rise
to multiple paths. As a consequence, a
self-averaging process takes place in (\ref{eq:csquaredlag}) 
and this prevents the indefinite growth of the energy. 
This is evident upon rewriting (\ref{eq:csquaredlag}) as
\begin{equation}
\fl \langle c^2({\bm x},t)\rangle =
\int_0^t  {\rm d}s \int \int
\langle P_2({\bm y}_1,{\bm y}_2,s|{\bm x},{\bm x},t)\rangle_v 
{\cal F}\left({|{\bm y}_1-{\bm y}_2|/ \ell_f}\right) {\rm d}{\bm y}_1
{\rm d}{\bm y}_2\,.
\label{eq:csquaredplag}
\end{equation}
The time integral is cut off at $|t-s| \gg 
{\cal T}_{\ell_f}$, that is for times larger than the
(finite) time needed by two coinciding particles to
separate by a distance larger than the forcing correlation length
$\ell_f$.
This is the mechanism leading to a finite dissipation of
energy. To summarize, the incompressibility and the roughness of the 
flow result in the 
dissipative anomaly through the explosive separation of particle paths.  
Further discussion on the role of
dissipative anomaly in passive scalar turbulence can be found in
\cite{K94,CKV97,Y97}.

\section{Two-dimensional turbulent convection}
\label{sec:2}
An interesting problem in the context of turbulent transport is the
advection of inhomogeneous temperature fields in a gravitational
field. Temperature fluctuations induce density fluctuations that in
turn, via buoyancy forces, affect the velocity field: hence the
temperature field is an active scalar \cite{MY75,S94}.  Here, we
consider two-dimensional convection, which is also of experimental
interest in Hele-Shaw flows \cite{bsq2dex}.  As an additional asset,
the two-dimensional problem is better suited for the study of scaling
properties since it allows to reach higher resolution and larger
statistics.

Two-dimensional convection is described by the following equations
\begin{eqnarray}
\partial_t a+ {\bm v}\cdot {\bm \nabla} a &=& \kappa \Delta\, a + f_a \;,
\label{eq:bsq1}\\
\partial_t {\bm v} + {\bm v}\cdot {\bm \nabla} {\bm v} &=&
-{\bm \nabla} p + \nu\Delta{\bm v} -\beta a {\bm g} -\alpha {\bm
v}\,,  \label{eq:bsq3}
\end{eqnarray}
where $a$ is the field of temperature fluctuations. The second
equation is the two-dimensional Navier-Stokes equation where ${\bm v}$
is forced by the buoyancy term $-\beta {\bm g}a$ in the Boussinesq
approximation \cite{MY75}; ${\bm g}=g \hat{\bm y}$ is the
gravitational acceleration and $\beta$ the thermal expansion
coefficient. Kinetic energy is removed at the large scales by the
friction term, $-\alpha {\bm v}$.  The friction is physically due to the
drag experienced by a thin (quasi $2d$) layer of fluid with the walls
or air \cite{S98,RW00}; $\alpha$ is related to the thickness of the
fluid layer. A passive scalar, $c$, evolving according to
Eq.~(\ref{eq:2}) in the same flow has been considered as well, for
comparison.

\begin{figure}[t!]
\begin{center}
\includegraphics[draft=false, scale=0.68, clip=true]{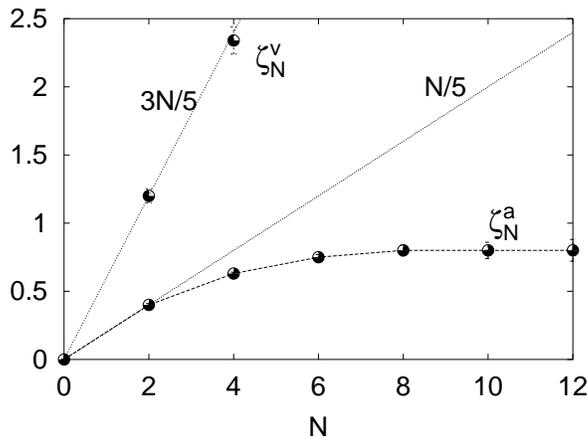} 
\caption{Scaling exponents of temperature, $\zeta^a_N$, and velocity,
$\zeta^v_N$. The straight lines are the dimensional predictions, $N/5$
for temperature, $3N/5$ for velocity. Details on the numerics may be
found in \protect{\cite{CMMV02}}.  Notice that at orders larger than
$N=8$ the temperature exponents saturate to a constant value
$\zeta^a_{\infty}\simeq 0.8$.}
\label{fig:bsq1}
\end{center}
\end{figure}

Before entering in the active/passive scalar issue, let us briefly
recall the phenomenology of 2D turbulent convection (for the 3D case
see, e.g., \cite{MY75,S94}). The balance of buoyancy and inertial terms
in (\ref{eq:bsq3}) introduces the Bolgiano lengthscale
$\ell_B$.  \cite{MY75}.  At small scales, $r\ll \ell_B$, the inertial
term is larger than buoyancy forces and the temperature is basically a
passive scalar.  At large scales, $r\gg \ell_B$, buoyancy dominates
and affects the velocity, which in two dimensions performs an inverse
energy cascade. However, at variance with the usual 2D Navier-Stokes
turbulence, the kinetic energy input rate $\varepsilon$ depends here on the
scale. Dimensional arguments yield $\varepsilon(r)=\beta {\bm g}\cdot
\langle {\bm v}({\bm x}+{\bm r},t)a({\bm x},t)\rangle \sim r^{4/5}$,
the Bolgiano scaling for the velocity structure functions
\begin{eqnarray}
S^v_N(r) \sim (\varepsilon(r) r)^{N/3}\sim r^{\zeta^v_N}\,,\qquad
\zeta^v_N=3N/5\,,
\end{eqnarray} 
and for temperature
\begin{eqnarray}
S^a_N(r) \sim r^{\zeta^a_N}\,,\qquad
\zeta^a_N=N/5\,.
\end{eqnarray} 
No intermittency corrections are reported for the velocity, 
whereas the temperature
field is strongly intermittent (see 
Fig.~\ref{fig:bsq1} and Refs.~\cite{CMV01,CMMV02}).  

Summarizing, the temperature fluctuations are injected at scales,
$\sim \ell_f$, pump kinetic energy through the buoyancy term, and a
non-intermittent velocity inverse cascade establishes with $\delta_r
v\sim r^{3/5}$. The presence of friction stabilizes the system
inducing a statistically steady state. The active and the passive
scalars are therefore 
transported by a self-similar, incompressible and rough flow. 
An intermittent cascade of fluctuations with
anomalous, universal scaling exponents, $\zeta^c_N$ is observed 
for the passive scalar.

\begin{figure}[t!]
\includegraphics[draft=false, scale=0.64, clip=true]{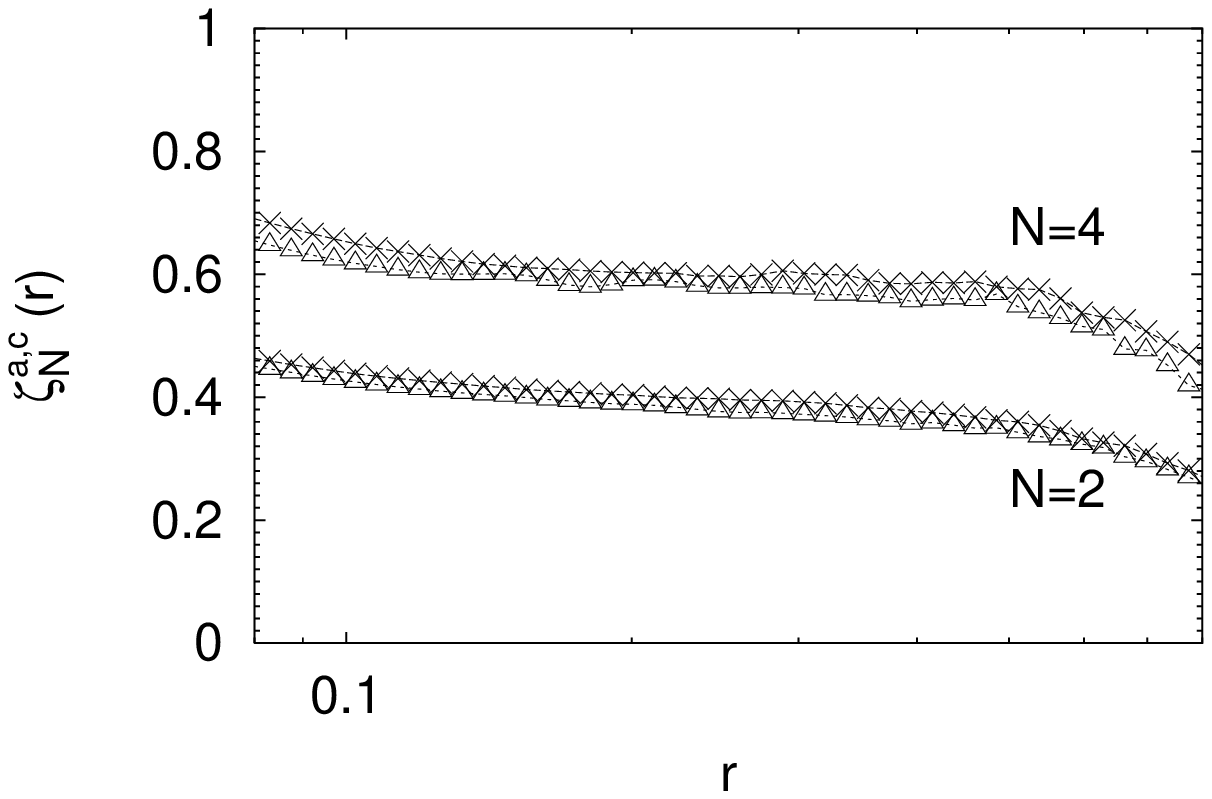} \includegraphics[draft=false, scale=0.64, clip=true]{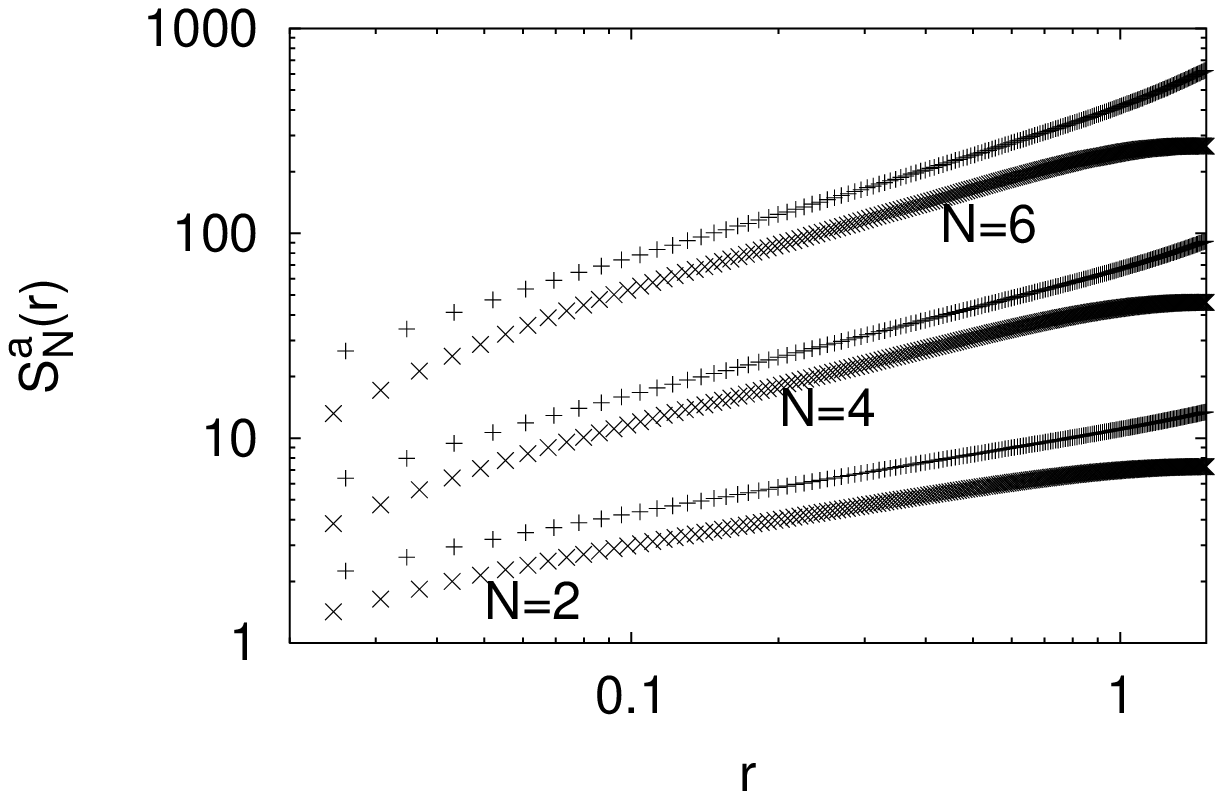} 
\caption{ (a) Local scaling exponents of temperature ($\times$) and
concentration ($\bigtriangleup$) fluctuations, $\zeta^{a,c}_N(r) = d
\ln S^{a,c}_N(r) / d \ln r$.  Temperature and concentration are driven
by independent Gaussian random forcing with correlation function as
Eq.~(\ref{eq:forcings}). (b) Temperature structure functions,
$S^a_N(r)$, for $N=2,4,6$, as a function of the separation $r$. The
two sets of curves are generated by using the random ($\times$) and
mean gradient ($+$) forcing. Note the parallelism within the inertial
range.  Curves have been multiplied for a numerical factor for
visualization purposes.}
\label{fig:bsq2}
\end{figure}

Our aim is to compare the statistical properties of the temperature
and the passive scalar fields. To this purpose, in Ref.~\cite{CMMV02},
Eqs.~(\ref{eq:bsq1}) and (\ref{eq:bsq3}) have been integrated with
$f_a$ and $f_c$, chosen as two independent realizations of a
stochastic, isotropic, homogeneous and Gaussian process of zero mean
and correlation:
\begin{equation}
\langle f_i({\bm x},t) f_j({\bm x}',t') \rangle= \delta_{ij}
 \delta(t-t'){\cal F}(|{\bm x}-{\bm x}'|/\ell_f)\,,
\label{eq:forcings}
\end{equation}
where ${\cal F}(r/\ell_f)=\exp(-r^2/(2\ell_f^{\;2}))$ decreases
rapidly as $r\geq \ell_f$. The labels are $i,j=a,c$.  The results of
this numerical study clearly confirm that temperature scaling
exponents are anomalous (Fig.~\ref{fig:bsq1}), and coincide with those
of the passive field: $\zeta^c_N=\zeta^a_N$ (Fig.~\ref{fig:bsq2}a). 

 In this system there is saturation of intermittency, i.e. for large
$N$ the scaling exponents saturate to a constant
$\zeta^{a,c}_{\infty}\approx 0.8$ (see Fig.~\ref{fig:bsq1}).  This
phenomenon, well known for passive scalars \cite{CLMV01}, is
physically related to the presence of abrupt changes in the spatial
structure in the scalar field (``fronts''). In the temperature field
these quasi-discontinuities correspond to the boundaries between hot
rising and cold descending patches of fluid \cite{CMV01}. It is worth
mentioning that saturation has been experimentally observed both for
passive scalars \cite{W00,MWAT01} and temperature fields
\cite{ZX02}. Clear evidences of saturation have recently been obtained
also in the convective atmospheric boundary layer exploiting the
large-eddy simulation technique \cite{AMR03}.

These findings point to the conclusion that the temperature and the
passive scalar have the same scaling laws. 
It remains to be ascertained whether the temperature scaling exponents are
universal with respect to the forcing. 
To this aim, a set of simulations has been performed 
in
Ref.~\cite{CMV01,CMMV02}, with a forcing that
mimics the effect of a superimposed mean gradient on the
transported temperature field
\begin{equation}
f_a({\bm x},t)= \gamma {\bm g}\cdot {\bm v}({\bm x},t)\,.
\label{eq:mg}
\end{equation}
Remarkably, the results show that the scaling exponents of 
the temperature field do not
depend on the injection mechanism (Fig.~\ref{fig:bsq2}b) suggesting 
universality \cite{CMMV02,CMV01}.  Another outcome of this
investigation is that the velocity field statistics itself
is universal with respect to the injection mechanism of the
temperature field. Indeed, ${\bm v}$ displays a close-to-Gaussian and
non-intermittent statistics with both forcing (\ref{eq:forcings}) and
(\ref{eq:mg}) \cite{CMMV02}.  This is most likely a consequence of the
observed universal Gaussian behavior of the inverse energy cascade in
two-dimensional Navier-Stokes turbulence \cite{PT98,BCV00}.  Indeed,
velocity fluctuations in two-dimensional convection also arise from an
inverse cascade process driven by buoyancy forces.

So far all the numerical evidences converge to the following global
picture of scaling and universality in two-dimensional turbulent
convection.  Velocity statistics is strongly universal with respect to
the temperature external driving, $f_a$.  Temperature statistics shows
anomalous scaling exponents that are universal and coincide with those
of a passive scalar evolving in the same flow. It is worth noticing
that similar findings have been obtained in the context of simplified
shell models for turbulent convection \cite{CCGP02,CCGP03}.

The observed universality of the temperature scaling exponents
suggests that a mechanism similar to that of passive scalars may be at
work, i.e.  that statistically preserved structures might exist also
for the (active) temperature.  Pursuing this line of thought, one may
be tempted to define them through the property $\frac{d}{dt}
\langle{\cal Z}_N^a\rangle_{\cal L}=0$ as for passive scalars (see
Eq.~(\ref{eq:sp})). However, statistically preserved structures are
determined by the statistics of particle trajectories which, through
the feedback of $a$ on ${\bm v}$, depend on $f_a$. Therefore, the
above definition does not automatically imply the universality of
${\cal Z}_N^a$, because Lagrangian paths depend on $f_a$. Nonetheless, the
observed universality of the statistics of ${\bm v}$ is sufficient to
guarantee the universality of the trajectories statistics, leading to
the conclusion that if ${\cal Z}_N^a$ exists it might be universal.
Since ${\cal Z}_N^c$ are also defined by the Lagrangian statistics
that is the same for $a$ and $c$, one might further conjecture that
${\cal Z}_N^a={\cal Z}_N^c$. This would explain the equality of
scaling exponents $\zeta^a_N=\zeta^c_N$.

It has to be remarked that this picture is not likely to be generic.
Two crucial points are needed to have the equality between active and
passive scalar exponents: {\it (i)\/} the velocity statistics should
be universal; {\it (ii)\/} the correlation between $f_a$ and the
particle paths should be negligible.  As we shall see in the
following, those two requirements are not generally met.

\section{Two-dimensional magnetohydrodynamics}
\label{sec:3}

\subsection{Direct and inverse cascades}
Magnetohydrodynamics (MHD) models are extensively used in the study of
magnetic fusion devices, industrial processing plasmas, and
ionospheric/astrophysical plasmas \cite{ZRS83}. MHD is the extension
of hydrodynamics to conductive fluids, including the effects of 
electromagnetic fields. When
the magnetic field, ${\bm b}$, has a strong large-scale component in one
direction, the dynamics is adequately described by the two-dimensional 
MHD equations \cite{B93}. Since the magnetic field ${\bm b}({\bm x},t)$ is
solenoidal, in $2d$ it can be represented in terms of the magnetic
scalar potential, $a({\bm x},t)$, i.e. ${\bm b}=-{\bm \nabla}^{\perp}
a=(-\partial_2 a,\partial_1 a)$.  The magnetic potential evolves
according to the advection-diffusion equation
\begin{equation}
\partial_t a+ {\bm v}\cdot {\bm \nabla} a = \kappa \Delta\, a + f_a \;,
\label{eq:mhd1}
\end{equation}
and will be our active scalar throughout this section. 
The advecting velocity field is driven by the Lorentz force,
$({\bm \nabla} \times {\bm b})\times {\bm b}=-\Delta a {\bm \nabla}a$,
so that the Navier-Stokes equation becomes
\begin{equation}
\partial_t {\bm v} + {\bm v}\cdot {\bm \nabla} {\bm v} = -{\bm \nabla}
p + \nu\Delta{\bm v} -\Delta a {\bm \nabla} a \,.
\label{eq:mhd2}
\end{equation}
The question is 
whether the picture drawn
for the temperature field in $2d$ convection applies to the
magnetic potential as well. 

\begin{figure}[t!]
\begin{center}
\includegraphics[draft=false, scale=0.68, clip=true]{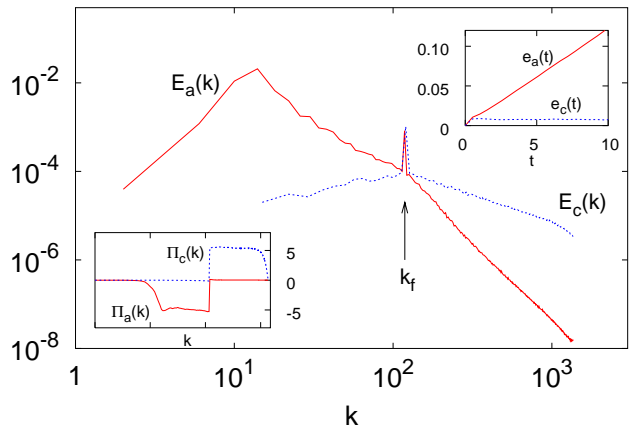} 
\caption{Power spectra of active (red) and passive (blue)
 scalar variances $E_a(k)=\pi k |\hat{a}({\bm k},t)|^2$ and
$E_c(k)=\pi k |\hat{c}({\bm k},t)|^2$.  In the lower left corner, the
fluxes of scalar variance $\Pi_{a,c}$ out of wavenumber $k$. Negative
values indicate an inverse cascade.  In the upper right corner, the
total scalar variance $e_{a,c}(t)= \int E_{a,c}(k,t)\,dk$.  The active
variance $e_a(t)$ grows linearly in time whereas $e_c(t)$ fluctuates
around a finite value (see text). The rate of active to passive scalar
dissipation is $\epsilon_a/\epsilon_c \simeq 0.005$.  All fields are
set to zero at $t=0$, and time is defined in units of eddy-turnover
time ${\cal T}=l_f/v_{\mathrm{rms}}$ where $l_f=2\pi/k_f$.  At $k<k_f$
we observe power-law behaviors $E_{a}(k) \sim k^{-2.0\pm 0.1}$ and
$E_c(k) \sim k^{0.7\pm 0.1}$, while at $k>k_f$ we find $E_a(k)\sim
k^{-3.6 \pm 0.1}$ and $E_c(k)\sim k^{-1.4\pm 0.1}$.}
\label{fig:mhd1} 
\end{center}
\end{figure}

Eqs.~(\ref{eq:mhd1}) and (\ref{eq:mhd2}) have two quadratic invariants
in the inviscid and unforced limit, namely the total energy ${1 \over
2}\int (v^2+b^2) {\rm d}{\bm x}$ and the mean square magnetic
potential ${1 \over 2}\int a^2 {\rm d}{\bm x}$. Using standard
quasi-equilibrium arguments \cite{B93}, an inverse cascade of magnetic
potential is expected to take place in the forced and dissipated case
\cite{Pouq}. This expectation has been confirmed in numerical
experiments \cite{Bis2}.  Let us now compare the magnetic potential
with a passive scalar evolving in the same flow.

We performed a high-resolution ($4096^2$ collocation points) direct
numerical simulations of Eqs.~(\ref{eq:mhd1})-(\ref{eq:mhd2}) along
with a passive scalar (\ref{eq:2}).  The scalar forcing terms
$f_a$ and $f_c$ are homogeneous independent Gaussian processes with
zero mean and correlation
\begin{equation}
\langle \hat{f}_i({\bm k},t)\hat{f}_j({\bm k}',t')\rangle
={F_0 \over (2\pi k_f)} \delta_{ij}\delta({\bm k}+{\bm
k}')\delta(k-k_f)\delta(t-t')
\label{eq:fband}
\end{equation}
where $i,j=a,c$. The injection length scale $l_f\sim 2\pi/k_f$ has
been chosen roughly in the middle of the available range of scales.
$F_0$ is the rate of scalar variance input.


In Fig.~\ref{fig:mhd1} we summarize the spectral properties of the two
scalars. The emerging picture is as follows.  While $a$ undergoes an
inverse cascade process, $c$ cascades downscale.  This striking
difference is reflected in the behavior of the dissipation.  The
active scalar dissipation, $\epsilon_a = \kappa |{\bm \nabla a}|^2$,
vanishes in the limit $\kappa\to 0$ -- no dissipative anomaly for the
field $a$. Consequently, the squared magnetic potential grows linearly
in time $e_a(t)={1\over 2} \int a^2 d{\bm x}\approx {1 \over 2} F_0
t$. On the contrary, for the passive scalar, a dissipative anomaly is
present and $\epsilon_c = \kappa |{\bm \nabla c}|^2$ equals the input
${1 \over 2} F_0$ holding $c$ in a statistically stationary
state. (See inset in Fig.~\ref{fig:mhd1}).

\begin{figure}[t!]
\begin{center}
\includegraphics[draft=false, scale=0.68, clip=true]{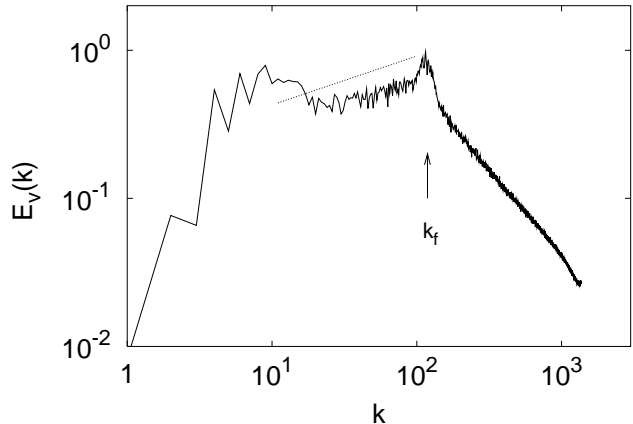} 
\caption{Velocity power spectrum $E_v(k)=\pi k |\hat{v}({\bm
k},t)|^2$.  For $k<k_f$, in agreement with previous simulations
\protect\cite{Bis2}, we observe $E_v(k)\sim k^{1/3}$ (dashed lines)
which deviates from the dimensional prediction $k^{-1/3}$.  In the
range $k>k_f$ a  scaling close to $k^{-5/3}$ is observed,
indicating that ${\bm v}$ is rough both in the inverse and in the
direct cascade range.}
\label{fig:mhd2}
\end{center} 
\end{figure}

The velocity field is rough (as confirmed by its spectrum, see
Fig.~\ref{fig:mhd2}) and incompressible, therefore particle paths are
not unique and explosively separate. This entails the
dissipative anomaly for passive scalars. 
On the contrary, the dissipative anomaly is absent for the magnetic potential,
in spite of the fact that the trajectories are the same -- the advecting 
velocity is the same. How can these two seemingly contradictory statements 
be reconciled ?

\subsection{Dissipative anomaly and particle paths}
The solution of the riddle resides in the relationship between
the Lagrangian trajectories and the active scalar input $f_a$.  These two
quantities are bridged by the Lorentz force appearing in
(\ref{eq:mhd2}).  To study the correlations between forcing and
particle paths we need to compute the evolution of the particle
propagator.  The relevant observables are the time sequences of
$\phi_{a,c}(s)=\int d{\bm y}\,f_{a,c}({\bm y},s) P({\bm y},s|{\bm
x},t)$. Indeed the equivalent of Eq.~(\ref{eq:plag}) can be written
for the active scalar as
\begin{equation}
a({\bm x},t)=\int_0^t ds\int d{\bm y}\,f_a({\bm y},s) 
P({\bm y},s|{\bm x},t)\;.
\label{eq:aplag}
\end{equation} 

The main difficulty encountered here  
is that $P$ evolves {\em backward} in time
according to (\ref{eq:Kolmo1}) and the condition $P({\bm y},t|{\bm
x},t)=\delta({\bm y}-{\bm x})$ is set at the final time $t$. Conversely,
the initial conditions on velocity and scalar fields are set at
the initial time.  The solution of this mixed initial/final value problem
is a non-trivial numerical task. To this aim, we
devised a fast and low-memory demanding algorithm to integrate
Eqs.~(\ref{eq:mhd1}) - (\ref{eq:mhd2}) and (\ref{eq:Kolmo1}) with 
the appropriate initial/final conditions. The details 
are given in Ref.~\cite{CCN03}.
\begin{figure}[t!]
\begin{center}
\includegraphics[draft=false, scale=0.6, clip=true]{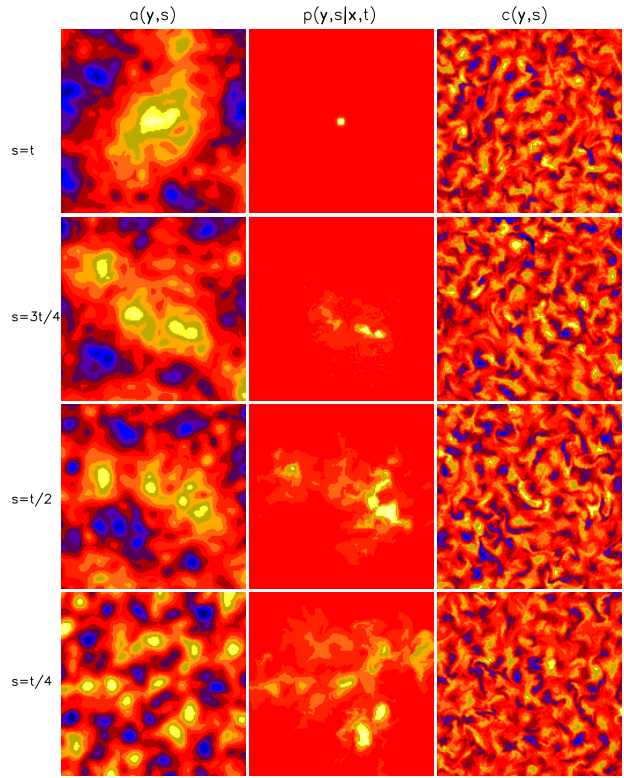} 
\caption{Time runs from bottom to top. First column: time evolution of
the active scalar field resulting from the numerical integration of
Eqs.~(\protect\ref{eq:mhd1}) and (\protect\ref{eq:mhd2}).  Second
column: backward evolution of the particle propagator according to
Eq.~(\protect\ref{eq:Kolmo1}).  Third column: time evolution of the
passive scalar field in the same flow.}
\label{fig:mhd3} 
\end{center}
\end{figure}

\begin{figure}[t!]
\begin{center}
\includegraphics[draft=false, scale=0.68, clip=true]{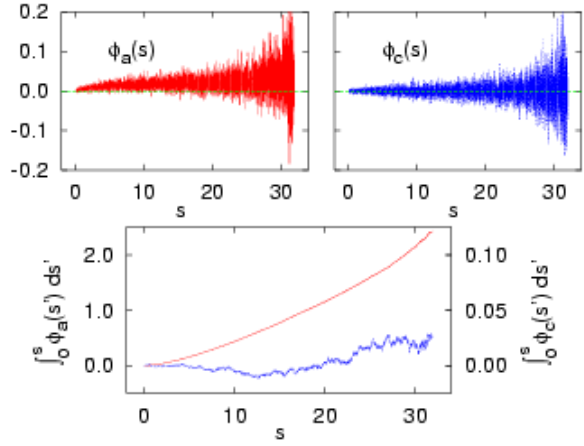} 
\caption{Top: $\phi_{a,c}(s)=\int d{\bm y}\,f_{a,c}({\bm y},s) P({\bm
y},s|{\bm x},t)$. The two graphs have the same scale on the vertical
axis. Here, $t=32$. Bottom: time integrals $\int_0^s \phi_{a}(s')\,ds'$
(upper curve) and $\int_0^s \phi_{c}(s')\,ds'$ (lower curve). Note
the different scale on the vertical axis. Recall that $\int_0^t
\phi_{a}(s')\,ds' =a({\bm x},t)$ (and similarly for $c$).  }
\label{fig:mhd4} 
\end{center}
\end{figure}

The typical evolution of the propagator is shown in the central column
of Fig.~\ref{fig:mhd3}. From its evolution we reconstructed the time
sequences of the forcing contributions $\phi_{a,c}(s)$ which,
integrated over $s$, give the amplitude of the scalar fields according
to (\ref{eq:plag}) and (\ref{eq:aplag}).  The time series
of $\phi_a(s)$ and $\phi_c(s)$ are markedly different
(Fig.~\ref{fig:mhd4}), the former being strongly skewed toward positive
values at all times. This signals that trajectories preferentially
select regions where $f_a$ has a positive sign, summing up
 forcing contributions to generate a typical variance of
$a$ of the order $F_0t$.  On the contrary, the passive scalar
sequence displays the usual features: $f_c$ is independent of
$P$ and their product can be positive or negative with 
equal probability on distant
trajectories. This ensures 
that the time integral in Eq.~(\ref{eq:plag}) averages
out to zero for $|s-t|>{\cal T}_{\ell_f}$ (see
Sect.~\ref{sec:lag}) and yields  $c^2 \sim F_0 {\cal
T}_{\ell_f}$.

As shown in the lower panel of Fig.~\ref{fig:mhd4} the effect of 
correlations between forcing and propagator is even more striking
comparing $\int_0^s {\rm d}s' \phi_a(s')$ with $\int_0^s {\rm d}s'
\phi_c(s')$. The conspicuous difference has to be related to
a strong spatial correlation between $P$ and $a$, as can be 
inferred from
\begin{equation}
\int_0^s \phi_a(s')\,ds'=\int d{\bm y}\, a({\bm y},s)P({\bm y},s|{\bm x},t)\,,
\label{eq:mhdcor}
\end{equation}
which can be derived from (\ref{eq:mhd1}) and (\ref{eq:Kolmo2}). An
equivalent relation  holds for $c$ as well.  
The comparison of the first and the
second column of Fig.~\ref{fig:mhd3} highlights the role of spatial 
correlations: the distribution of particles follows the distribution of
the active scalar. This amounts to saying that large-scale scalar
structures are built out of smaller ones that coalesce together
\cite{Bis2}.  This has to be contrasted with the absence of
large-scale correlations between the propagator and the passive scalar
field (second and third column of Fig.~\ref{fig:mhd3}).

Let us now clarify the mechanism for the absence of dissipative
anomaly.  Consider the squared active field $a^2$. It can be
expressed in two equivalent ways. On one hand, it can be written as the
square of (\ref{eq:aplag}). On the other hand, multiplying
(\ref{eq:mhd1}) by $2 a$ one obtains the equation
\begin{equation}
\partial_t a^2 +{\bm v}\cdot\nabla a^2 = \kappa \Delta a^2 +2af_a
-2\epsilon_a\,.
\label{eq:a2}
\end{equation}
Exploiting the absence of dissipative anomaly,
$\epsilon_a=0$, Eq.~(\ref{eq:a2}) 
reduces to a transport equation that can be solved in
terms of particle trajectories. The comparison of the two 
expressions yields \cite{nota1}
\begin{eqnarray}
\int_0^t\hspace{-3pt} ds \hspace{-2pt}\int_0^t \hspace{-3pt}ds'
\hspace{-3pt}\int\hspace{-8pt}\int\hspace{-3pt}
f_a({\bm y},s) f_a({\bm y}',s')
P({\bm y},s|{\bm x},t)P({\bm y}',s'|{\bm x},t)&=& \nonumber\\
\int_0^t\hspace{-3pt} ds \hspace{-2pt}\int_0^t \hspace{-3pt}ds'
\hspace{-3pt}\int\hspace{-8pt}\int\hspace{-3pt}
f_a({\bm y},s) f_a({\bm y}',s')
P({\bm y}',s';{\bm y},s|{\bm x},t)\,,& &
\label{eq:mhd7}
\end{eqnarray}
where $P({\bm y}',s';{\bm y},s|{\bm x},t)= P({\bm y},s|{\bm
x},t)P({\bm y}',s'|{\bm y},s)$ denotes the probability that a
trajectory ending in $({\bm x},t)$ were in $({\bm y},s)$ and $({\bm
y}',s')$.  Integration over ${\bm y}$ and ${\bm y}'$ is implied.  
Eq.~(\ref{eq:mhd7}) amounts to saying that 
\begin{equation}
\left\langle \int_0^t f_a({\bm \rho}(s),s)ds \right\rangle_{\bm w}^2\!\!=\! 
\left\langle \left[\int_0^t f_a({\bm \rho}(s),s) ds\right]^2
\right\rangle_{\bm w}\,,
\label{eq:mhd8}
\end{equation}
meaning that $\int_0^t f_a({\bm \rho}(s),s)ds$ is a {\em non-random}
variable over the ensemble of trajectories.  The above procedure can
be generalized to show that $\langle \int_0^t f_a({\bm \rho}(s),s)ds
\rangle_{\bm w}^N= \langle [\int_0^t f_a({\bm \rho}(s),s) ds]^N
\rangle_{\bm w}$.

In plain words, the absence of the dissipative anomaly is equivalent
to the property that along any of the infinite trajectories ${\bm
\rho}(s)$ ending in $({\bm x},t)$ the quantity $\int_0^t f_a({\bm
\rho}(s),s)ds$ is exactly the same, and equals $a({\bm
x},t)$. Therefore, a single trajectory suffices to obtain the value of
$a({\bm x},t)$, contrary to the passive case where different
trajectories contribute disparate values of $\int_0^t f_c({\bm
\rho}(s),s)ds$, with a typical spread $\epsilon_c t$, and only the
average over all trajectories yields the correct value of $c({\bm
x},t)$.  In the unforced case particles move along isoscalar lines:
this is how non-uniqueness and explosive separation of trajectories is
reconciled with the absence of dissipative anomaly.

Inverse cascades appear also for passive scalars in compressible flows
\cite{GV00}.  There, the ensemble of the trajectories collapses onto a
unique path, fulfilling in the simplest way the
constraint~(\ref{eq:mhd7}). In MHD the constraint is satisfied thanks
to the subtle correlation between forcing and trajectories peculiar to
the active case.

Magnetohydrodynamics in two dimension represents an ``extreme''
example of the effect of correlations among Lagrangian paths and the
active scalar input. The property that all trajectories ending in the same
point should contribute the same value of the input poses a global
constraint over the possible paths.  

Before discussing the statistical properties of $a$, on the basis of
the previous discussion it is instructive to reconsider the concept of
dissipative anomaly in general scalar turbulence.

\subsection{Dissipative anomaly  revisited}
In this subsection we give an alternative interpretation of
dissipative anomaly. To this aim, let us denote with $\Theta({\bm
x},t)$ a generic scalar field, regardless of its passive or active
character.  The scalar evolves according to the transport equation
\begin{equation}
\partial_t \Theta+{\bm v}\cdot {\bm \nabla}\Theta=\kappa \Delta \Theta
+f_{\Theta}\,.
\label{eq:ad}
\end{equation}
We can formally solve Eq.~(\ref{eq:ad}) by the method of characteristics,
i.e. in terms of the stochastic ordinary differential equations 
\begin{eqnarray}
{{\rm d} {\bm \rho}(t) \over {\rm d}t} &=& {\bm v}({\bm
\rho}(t),t)+\sqrt{2\kappa} \dot{\bm w}, \label{eq:paths1}\\ {{\rm
d}\vartheta(t) \over {\rm d}t} &=& f_{\Theta} ({\bm \rho}(t),t)\,.
\label{eq:paths2}
\end{eqnarray}
At variance with the procedure adopted in Sect.~\ref{sec:lag} 
here we are not conditioning {\it a
priori} the paths to their final positions.  The Eulerian value of the
field is recovered once the average over the all paths, ${\bm \rho}$,
landing in $({\bm x},t)$ is performed, i.e. $\Theta ({\bm
x},t)=\langle \vartheta(t) \rangle_{\bm w}$. Recall that if the flow
is non-Lipschitz continuous, such paths do not collapse onto a single
one also for $\kappa \to 0$.

We can now define $P({\bm x}, \vartheta,t| {\bm x}_0,\vartheta_0,0)$
as the probability that a path which started in ${\bm x}_0$ at time
$0$ with $\vartheta_0=\Theta({\bm x}_0,0)$ and arrives in ${\bm x}$ at
time $t$ carrying a scalar value $\vartheta$. Notice that the
conditioning is now on the initial value, and $P$ evolves according to
the Kolmogorov equation:
\begin{equation}
\partial_t P +{\bm v}\cdot {\bm \nabla}_{\bm x}P+f_{\Theta} \nabla_{\vartheta} P=
\kappa \Delta P
\label{eq:fkpl}
\end{equation}
with initial condition $P({\bm x}, \vartheta,t|{\bm x}_0,\vartheta_0,0)=
\delta({\bm x}-{\bm x}_0)\delta(\vartheta-\vartheta_0)$ and
$\vartheta_0=\Theta({\bm x},0)$. 
\begin{figure}[t!]
\begin{center}
\includegraphics[draft=false, scale=0.68, clip=true]{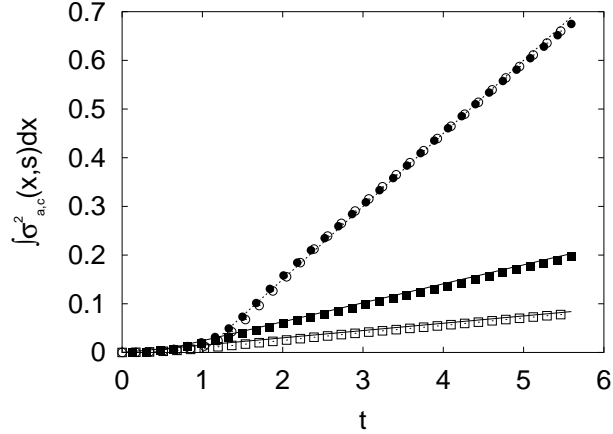} 
\caption{$\int {\rm d}{\bm x} \, \sigma^2_a({\bm x},t)$ (boxes) and
$\int {\rm d}{\bm x} \, \sigma^2_c({\bm x},t)$ (circles) vs time.  We
have integrated Eqs.~(\ref{eq:mhd1}) and (\ref{eq:2}) for two
different values of the diffusivity $\kappa=0.003$ (filled symbols) at
resolution $512^2$ and $\kappa=0.001$ (open symbols) at resolution
$1024^2$. The straight lines indicates the growth laws $2\epsilon_a t$
and $2\epsilon_c t$ for the two values of $\kappa$.  Note that
$\epsilon_c$ does not depend on $\kappa$ consistently with the
presence of dissipative anomaly, while $\epsilon_a$ decreases as
$\kappa$ decreases. The variance has been evaluated averaging over
$10^6$ Lagrangian paths evolving according to (\ref{eq:paths1}).  The
Lagrangian scalar values $\vartheta_a$ and $\vartheta_c$ (where
$a({\bm x},t)=\int \vartheta_a {\cal P}({\bm x},\vartheta_a,t) {\rm d}
\vartheta_a$, the equivalent relation holds between $c$ and
$\vartheta_c$) have been computed integrating (\ref{eq:paths2}) both for the
active $f_a$ and passive $f_c$ forcings along
each paths. The forcings are chosen as in (\ref{eq:fband}) 
The initial Eulerian $a({\bm x},0)$, $c({\bm x},0)$ and
Lagrangian $\vartheta_a(0)$, $\vartheta_c(0)$ fields have been set to
zero. Times is measured in eddy turnover times.}
\label{fig:mhd5} 
\end{center}
\end{figure}

Integrating over the initial conditions we define now
the probability density ${\cal P}({\bm
x},\vartheta,t) = \int P({\bm x}, \vartheta,t|{\bm x}_0,\Theta({\bm
x}_0,0),0){\rm d}{\bm x}_0$.  ${\cal P}$ still obeys to
(\ref{eq:fkpl}) with initial condition ${\cal P}({\bm
x},\vartheta,0)\!=\!\delta(\vartheta-\Theta({\bm x}_0,0))$, and represents
the probability that a path arrives in $({\bm x},t)$ carrying a scalar
value $\vartheta(t)\!=\!\int_0^t f_{\Theta}(\rho(s),s){\rm d}s$.

Let us now look at the variance of the distribution of such values,
i.e.  $\sigma^2_{\Theta}({\bm x},t)\!=\!\int \vartheta^2 {\cal P} {\rm
d}\vartheta- (\int \vartheta {\cal P} {\rm d}\vartheta)^2$. From
Eq.~(\ref{eq:fkpl}) it is easy to derive the following equation
\begin{equation}
\partial_t \sigma^2_{\Theta}({\bm x},t)\!+\!{\bm v}\cdot\!{\bm \nabla}_{\bm x}
\sigma^2_{\Theta}({\bm x},t)\!=\!\kappa \Delta \sigma^2_{\Theta}({\bm x},t)
\!+\!2\epsilon_{\Theta}({\bm x},t)
\end{equation}
where  $\epsilon_{\Theta}({\bm x},t)=\kappa |{\bm
\nabla}_{\bm x} \int \vartheta {\cal P} ({\bm x},\vartheta,t) {\rm d}
\vartheta|^2$. In the Eulerian frame
$\epsilon_\Theta({\bm x},t)=\kappa |{\bf \nabla} \Theta({\bm x},t)|^2$,
i.e. the local dissipation field.
\begin{figure}[t!]
\includegraphics[draft=false, scale=0.65, clip=true]{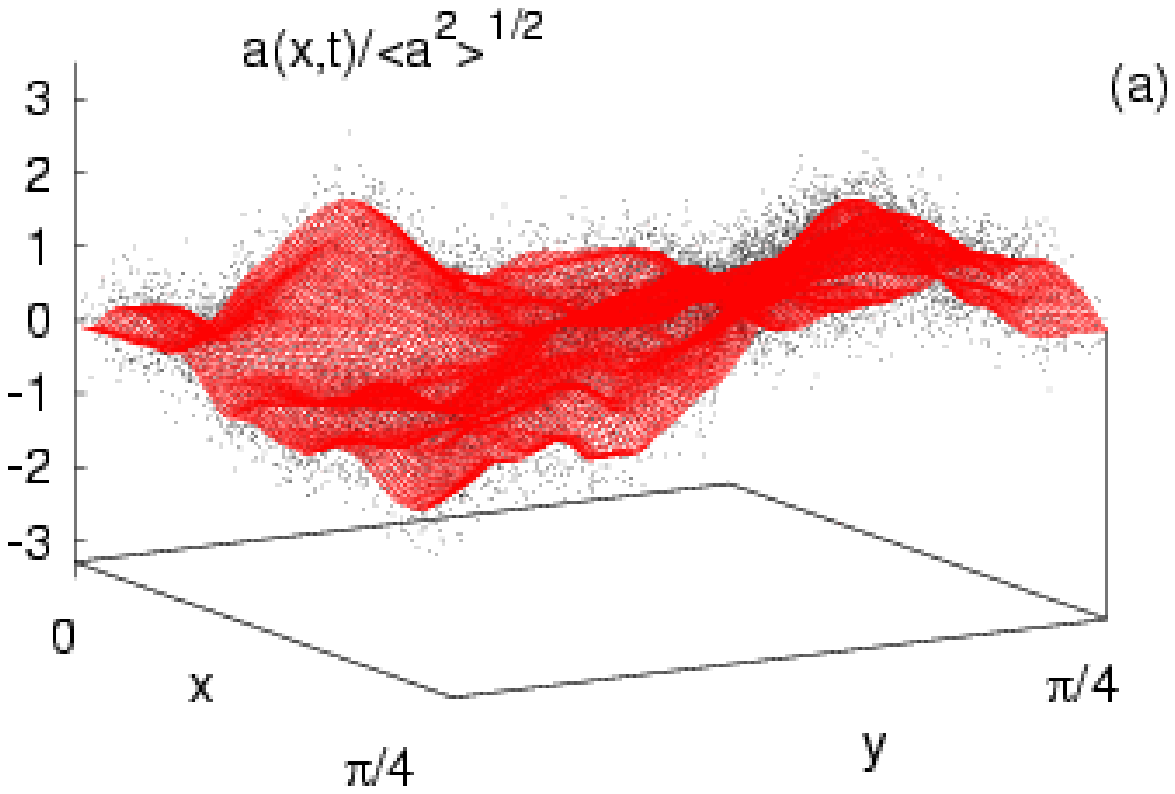} 
\includegraphics[draft=false, scale=0.65, clip=true]{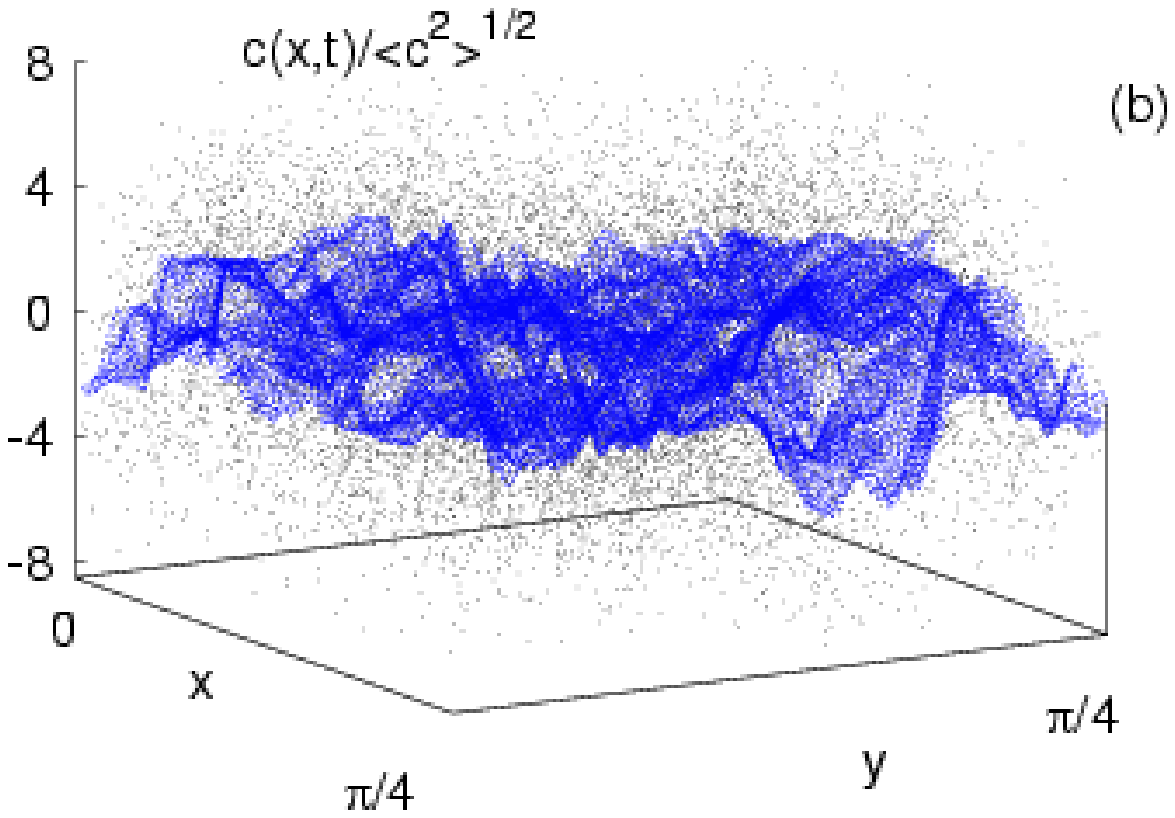} 
\caption{(a) The surface is the instantaneous Eulerian magnetic potential field
$a({\bm x},t)$. The dots represent the positions of
particles in the $({\bm x},\vartheta_a)$-space at time $t$. (b) The
same for the passive scalar, $c$. The time $t$
corresponds to the largest time in Fig.~\protect\ref{fig:mhd5}. For
visualization purposes only the portion $[0,\pi/4]\times [0,\pi/4]$ is
displayed.  In (a) the cloud of dots closely follows the the
magnetic potential surface, while in (b) they are considerably more 
dispersed (notice the
differences in the values of the $\vartheta$-axis).} \label{fig:mhd6} 
\end{figure}

Integrating over ${\bm x}$, we end up with
\begin{equation}
{{\rm d}\over {\rm d}t} \int \sigma^2_{\Theta}({\bm x},t) {\rm d}{\bm x}=
2\int \epsilon_{\Theta}({\bm x},t) {\rm d}{\bm x}=2\epsilon_{\Theta}\,,
\end{equation}
where $\epsilon_{\Theta}=\langle \kappa |{\bf \nabla} \Theta({\bm
x},t)|^2\rangle$ is the average dissipation rate of
$\langle\Theta^2\rangle/2$.  Therefore, if $\Theta^2$ cascades toward
the small scales with a finite (even in the limit $\kappa \to 0$)
dissipation, $\epsilon_{\Theta}$, the variance of the distribution of
the values of $\vartheta(t)=\int_0^t {\rm d}s f_{\Theta}({\bm
\rho}(s),s)$ will grow linearly in time. Conversely, for
an inverse cascade of $\Theta^2$ in the absence of dissipative anomaly
$\int {\rm d}{\bm x} \sigma^2_{\Theta}({\bm x},t)=0$, corresponding to a
singular distribution ${\cal P}({\bm x},\vartheta,t)=
\delta(\vartheta-\Theta({\bm x},t))$.

In Fig.~\ref{fig:mhd5} we show the time evolution of $\int {\rm d}{\bm
x} \, \sigma^2_{a,c}({\bm x},t)$ in the case of $2d$ MHD which
confirms the above findings.

The fact that, for inverse cascading scalars, the probability density
${\cal P}$ collapses onto a $\delta$-function in the limit of
vanishing diffusivity amounts to saying that particles in the $({\bm
x, \vartheta})$ space do not disperse in the $\vartheta$-direction but
move remaining attached to the surface $\vartheta=\Theta({\bm x},t)$.
(see Fig.~\ref{fig:mhd6}a). This is related to the strong spatial
correlations between $a$ and the Lagrangian propagator observed in
Fig.~\ref{fig:mhd3}.  On the contrary, for a direct cascade of scalar
such correlations do not exist and dissipation takes place because of
dispersion in the $\vartheta$-direction (see Fig.~\ref{fig:mhd6}b).

\subsection{Eulerian statistics}

\subsubsection{Single-point statistics}
\begin{figure}[t!]
\begin{center}
\includegraphics[draft=false, scale=0.68, clip=true]{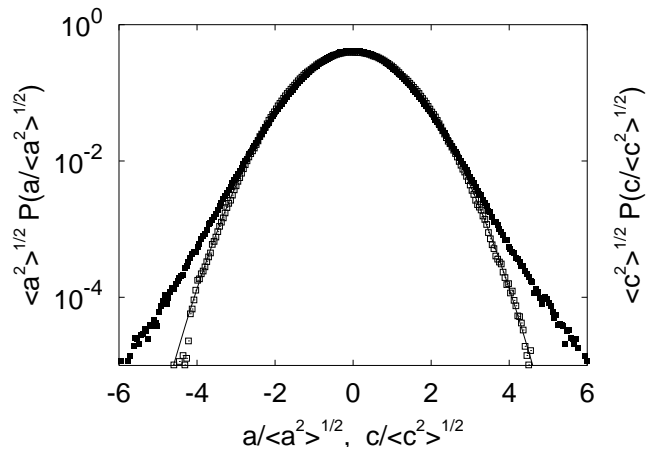} 
\caption{Pdf's of active (empty boxes) and passive (full boxes)
scalar fields normalized by their standard deviation.  The active
scalar pdf is indistinguishable from a Gaussian (solid line).}
\label{fig:mhd7} 
\end{center}
\end{figure}

The statistical properties of the magnetic potential $a$ and the
passive scalar $c$ are markedly different already at the level of the
single-point statistics. The pdf of $a$ is Gaussian, with zero mean
and variance $F_0 t$ (Fig.~\ref{fig:mhd7}).  Conversely, the pdf of
$c$ is stationary and supergaussian (see Fig.~\ref{fig:mhd7}),
as it generically happens for passive fields sustained by a Gaussian
forcing in rough flows \cite{FGV01}.

The Gaussianity of the pdf of $a$ is a straightforward consequence of
the vanishing of active scalar dissipation. This is simply derived by
multiplying Eq.~(\ref{eq:mhd1}) for $2n a^{2n-1}$ and averaging over
the forcing statistics. The active scalar
moments obey the equation $\partial_t \langle a^{2n} \rangle
= n(2n-1)F_0 \langle a^{2n-2} \rangle$ (odd moments vanish by
symmetry) whose solutions are the Gaussian moments: $\langle a^{2n}
\rangle = (2n-1)!!(F_0t)^n$.
An equivalent derivation can be obtained in Lagrangian terms.
Following the same steps which lead from Eq.~(\ref{eq:a2}) to (\ref{eq:mhd7})
it is easy to derive the following expression 
\begin{equation}
a^{2n}({\bm x},t)\!=\! 2n\! \!\!\int_0^t\!\! {\rm d}s_1 \!\! 
\int \!\! {\rm d}{\bm y}_1
P({\bm y}_1,s_1|{\bm x},t)
f_a({\bm y}_1,s_1) a^{2n-1}({\bm y}_1,s_1)\,,
\label{eq:a^n}
\end{equation}
which after integrating over ${\bm x}$ and averaging over the forcing
statistics reduces to
\begin{equation}
\langle a^{2n}\rangle(t)= n(2n\!-\!1) F_0 \int_0^t {\rm d}s \langle
a^{2n\!-\!2} \rangle(s)\,,
\label{eq:a^nmed}
\end{equation}
which, unraveling the hierarchy, yields Gaussian moments 
written above.
In passing from Eq.~(\ref{eq:a^n}) to (\ref{eq:a^nmed}) we used 
the property that $\int {\rm d}{\bm x} P({\bm y},s|{\bm x},t)=1$
(which is ensured by Eq.~(\ref{eq:Kolmo1})), and Gaussian integration
by parts.

\subsubsection{Multi-point statistics and the 
absence of anomalous scaling in the inverse cascade}
\label{sec:mhd2pts}
\begin{figure}[t!]
\includegraphics[draft=false, scale=0.68, clip=true]{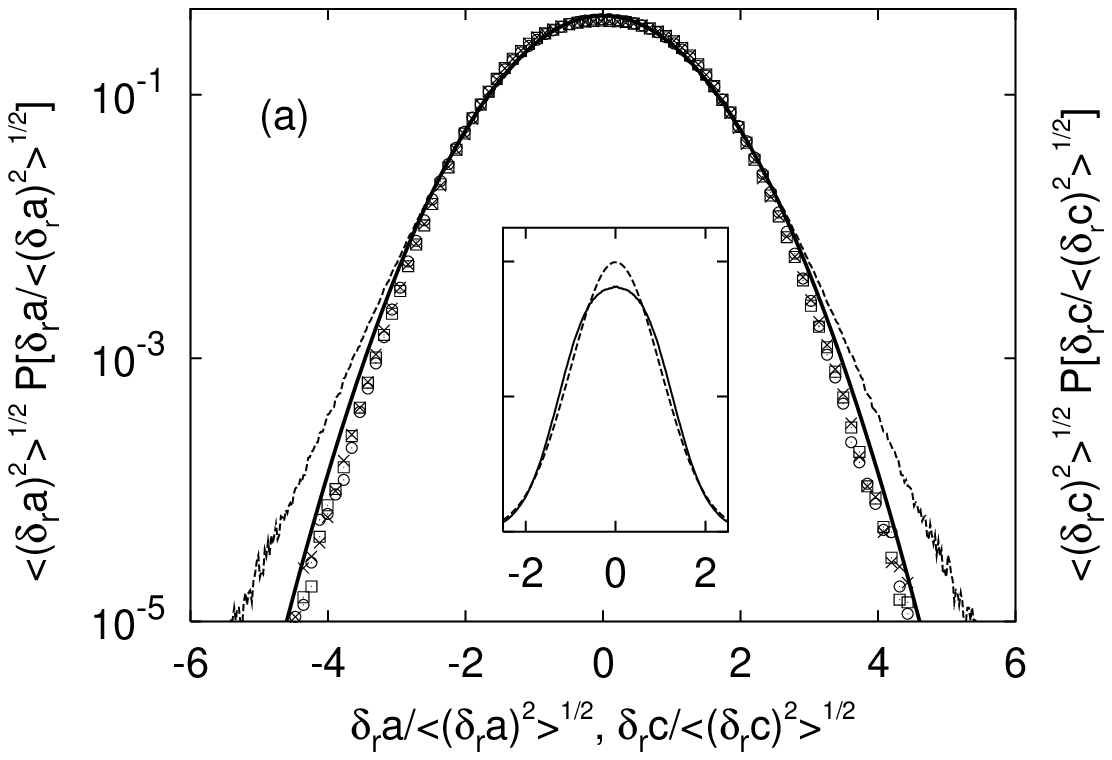} 
\includegraphics[draft=false, scale=0.68, clip=true]{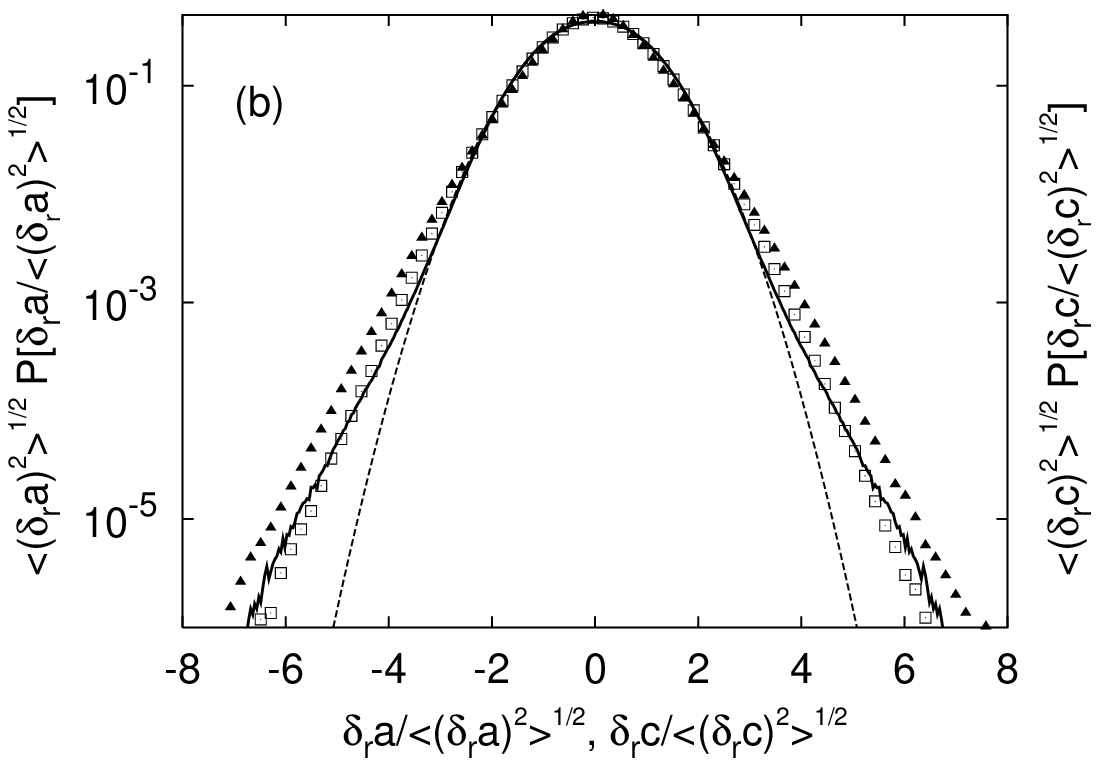} 
\caption{ (a) Rescaled pdf of the normalized active scalar increments,
$\delta_r a/\langle(\delta_r a)^2 \rangle^{1/2}$, for three
separations in the $r>\ell_f$ range: $r=2\ell_f$ ($\times$) ,$3\ell_f$
($\circ$), $4\ell_f$ ($\Box$).  The normalized passive scalar
increments pdf at the same scales (dashed line) and the Gaussian 
(solid line) are shown for comparison . In the inset the rescaled
pdf of $\delta_r a$ for $r=3\ell_f$ (solid line) is shown in the interval
$[-2.5:2.5]$ in linear scale, to emphasize the deviation from a
Gaussian (dashed line). The flatness is $\approx 2.76$, significantly
smaller than the Gaussian value $3$.  (b) Rescaled pdf of the
passive scalar increments, $\delta_r c/\langle(\delta_r
c)^2 \rangle^{1/2}$, for two separations in the $r<\ell_f$
range:  $r=0.3\ell_f$ (full triangles) and $0.6\ell_f$
($\Box$).  The rescaled pdfs of active scalar increments at the same
scales (solid line) and a Gaussian (dashed line) are reported for
comparison. }
\label{fig:mhd8} 
\end{figure}

The results about 
two-point statistics of the magnetic potential and passive scalar
fields is summarized in Fig.~\ref{fig:mhd8}. 
In the inverse cascade scaling range $r>\ell_f$ the rescaled
pdf of $\delta_r a$ at different values of the separation $r$ collapse
onto the same curve indicating {\it absence of anomalous scaling}
(Fig.~\ref{fig:mhd8}a).  By contrast, in the scaling range $r<\ell_f$
the rescaled pdfs of passive scalar increments at different $r$ do not
collapse (Fig.~\ref{fig:mhd8}b), i.e. we have anomalous scaling, as
expected.

\begin{figure}[t!]
\begin{center}
\includegraphics[draft=false, scale=0.68, clip=true]{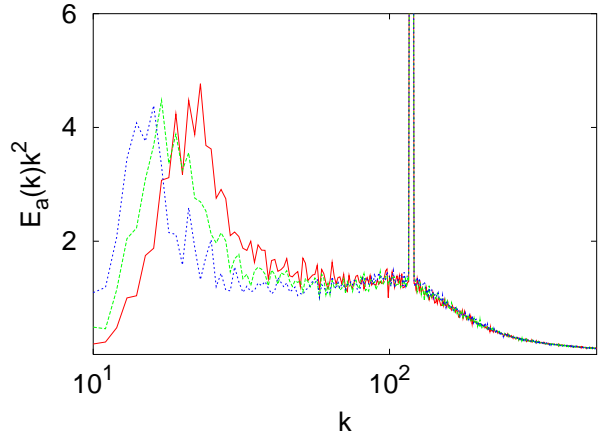} 
\caption{Compensated power spectra of active scalar variances
$E_a(k)k^2$ at three different times of the evolution. Note that
an increasingly clear plateau at $k<k_f$ appears.}
\label{fig:mhd10} 
\end{center}
\end{figure}

Let us now consider in quantitative terms the scaling behaviors.
Since in our simulations the scaling range at $r<\ell_f$ ($k>k_f$) is
poorly resolved, we do not enter the much debated issue of the
scaling of the magnetic  and velocity fields in this
range of scales (see, e.g., \cite{Bis2,Pouq,Srid,Verm,Pouq2,Bis3}).
We just mention that current opinions are divided between the
Kolmogorov scaling $\delta_r v \sim r^{1/3}$, and the
Iroshnikov-Kraichnan $\delta_r v \sim r^{1/4}$, corresponding to
spectral behaviors such as $E_v(k)\sim k^{-5/3}$ and $E_v(k)\sim
k^{-3/2}$, respectively.  Both theories agree on the smooth
scaling behavior for the magnetic potential, $\delta_r a\sim r$, that
is observed numerically (see Fig.~\ref{fig:mhd1}).
In the range of scales $r>\ell_f$ ($k<k_f$) standard dimensional
arguments predict $E_a(k) \sim k^{-7/3}$ \cite{B93,Pouq,Bis2}, which
is different from our findings $E_a(k)\sim k^{-2}$
(Fig.~\ref{fig:mhd10}). In real space this means that $\delta_r a\sim
r^{1/2}$, which is dimensionally compatible with scaling behavior
$\delta_r v = [{\bm v}({\bm x}+{\bm r},t)-{\bm v}({\bm x},t)]\cdot
{\hat{\bm r}} \sim r^0$ for $r > \ell_f$ (as suggested by the velocity
spectrum, see Fig.~\ref{fig:mhd2}), and the Yaglom relation $\langle
\delta_r v (\delta_r a)^2 \rangle \simeq F_0 r$ \cite{MY75}. As a side
remark, note that the argument for $E_a(k)\sim k^{-7/3}$ rests on the
assumption of locality for velocity increments, a hypothesis
incompatible with the observed velocity spectrum at $k<k_f$ (see
Fig.~\ref{fig:mhd2}).

 It is worth remarking that the increments of active scalar $\delta_r
a= a({\bm x}+{\bm r},t)-a({\bm x},t)$ eventually reach a stationary
state in spite of the growth of $a^2$.
The distribution of of $\delta_r a$
is subgaussian (Fig.~\ref{fig:mhd8}). This behaviour can be 
explained recalling that
$\delta_r a$ is the difference of two Gaussian variables, which are
however strongly correlated (indeed the main contributions will come
from ${\bm x}+{\bm r}$ and ${\bm x}$ inside the same island, see
Fig.~\ref{fig:mhd3}). This correlation leads to cancellations
resulting in the observed subgaussian probability density function.

Let us now focus on the most interesting aspect, namely the absence of
anomalous scaling in the inverse cascade range. The absence of
intermittency seems to be a common feature of inverse cascading
systems as passive scalars in compressible flows \cite{GV00,CKV97} and
the velocity field in two dimensional Navier Stokes turbulence
\cite{BCV00}. This leads to the conjecture that an universal mechanism
may be responsible for the self similarity of inverse cascading
systems. While $2d$ Navier-Stokes turbulence is still far to be
understood, the absence of anomalous scaling in passive scalars
evolving in compressible flows has been recently understood in terms
of the collapse of the Lagrangian trajectories onto a unique path
\cite{CKV97,GV00}. Briefly, the collapse of trajectories
allows to express the $2N$-order structure function in terms of
 two-particle propagators, $\langle P_2\rangle_v$, instead of 
the $2N$-particle propagator, $\langle P_{2N}\rangle_v$.  While the
latter may be dominated by a zero mode with a non trivial anomalous
scaling, the former are not anomalous and lead to dimensional
scaling.
 
The above argument cannot be simply exported to the
magnetic potential inverse cascade: first,
the Lagrangian paths do not collapse onto a unique one; second, the
correlation between the $f_a$ and ${\bm v}$ does not allow to split
the averages. 
However, the former difficulty 
can be overcome. Indeed the property that all paths landing in the same point 
contribute the same value has important consequences
also in the multipoints statistics.  
Proceeding as for the derivation of Eq.~(\ref{eq:mhd7}), it is possible 
to show that $m$
trajectories are enough to calculate the product of arbitrary powers
of $a$ at $m$ different points $\langle a^{n_1}({\bm x}_1,t)\dots
a^{n_m}({\bm x}_m,t)\rangle$.  In particular, structure functions
$S^{a}_N(r)=\langle (\delta_r a)^N \rangle$ for any order $N$ involve
only two trajectories.  This should be contrasted with the passive scalar,
where the number of trajectories increases with $N$ and this is at the
core of anomalous scaling of passive fields \cite{FGV01}.

\section{Two-dimensional Ekman turbulence}
\label{sec:4}

Let us now consider the case of scalars that act on the velocity field
through a functional dependence.  Among them, probably the best known 
case is vorticity in two dimensions  $a={\bm \nabla}\times {\bm v}$.
It obeys the Ekman-Navier-Stokes equation
\begin{equation}
\partial_t a+{\bm v}\cdot{\bm \nabla}a=\kappa \Delta a-\alpha \,a+f_a\,,
\label{eq:eck1}
\end{equation}
The term  $-\alpha \,a$ models
the Ekman drag experienced by
a thin fluid layer with the walls or the surrounding air \cite{S98,RW00}. 

In the absence of friction ($\alpha=0$) dimensional arguments
\cite{kraichnan67,batchelor69}, confirmed by experiments \cite{PT98,
PJT99} and numerical simulations \cite{BCV00}, give the following
scenario.  At large scales an inverse (non-intermittent) cascade of
kinetic energy takes place with $E_v(k)\sim k^{-5/3}$ and $E_a(k)\sim
k^{1/3}$.  At small scales the enstrophy, $\langle a^2 \rangle$,
performs a forward cascade with $E_a(k)\sim k^{-1}$ and $E_v(k)\sim
k^{-3}$, meaning that the velocity field is smooth in this range of
scales.

\begin{figure}[t!]
\hspace{-.5cm}\includegraphics[draft=false, scale=0.65, clip=true]{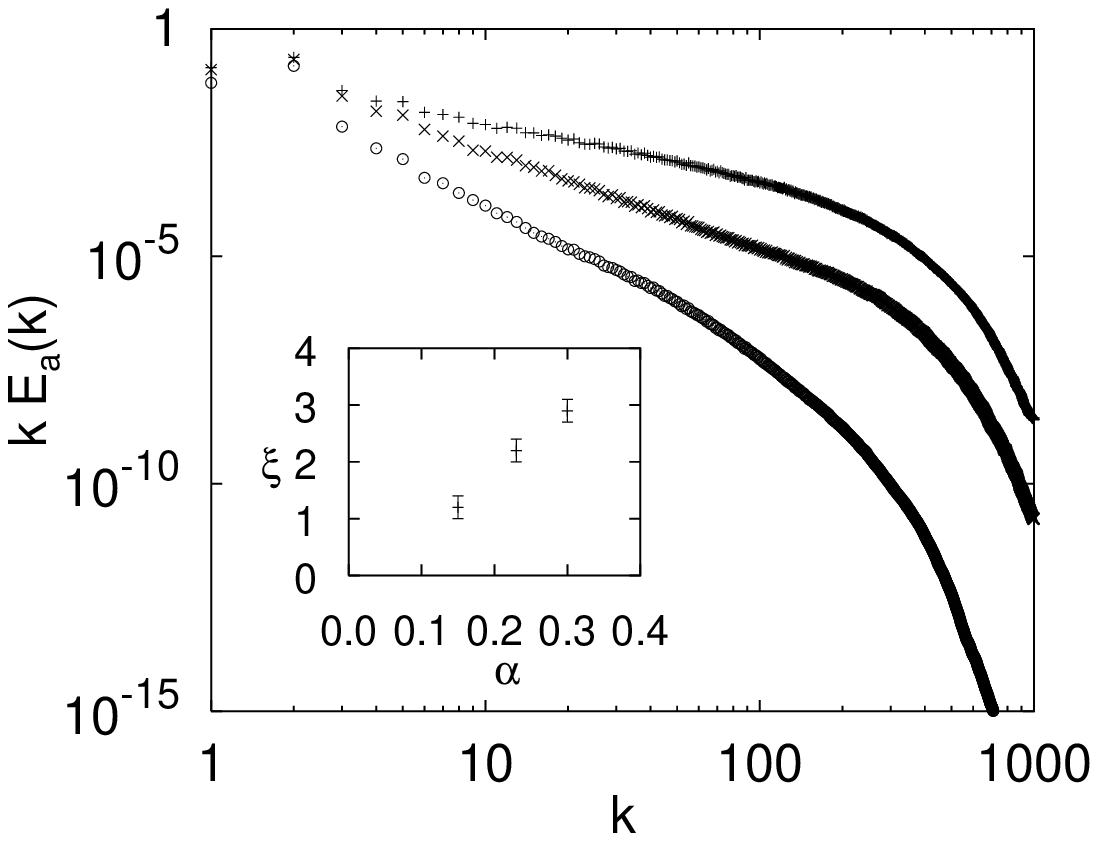} 
\includegraphics[draft=false, scale=0.65, clip=true]{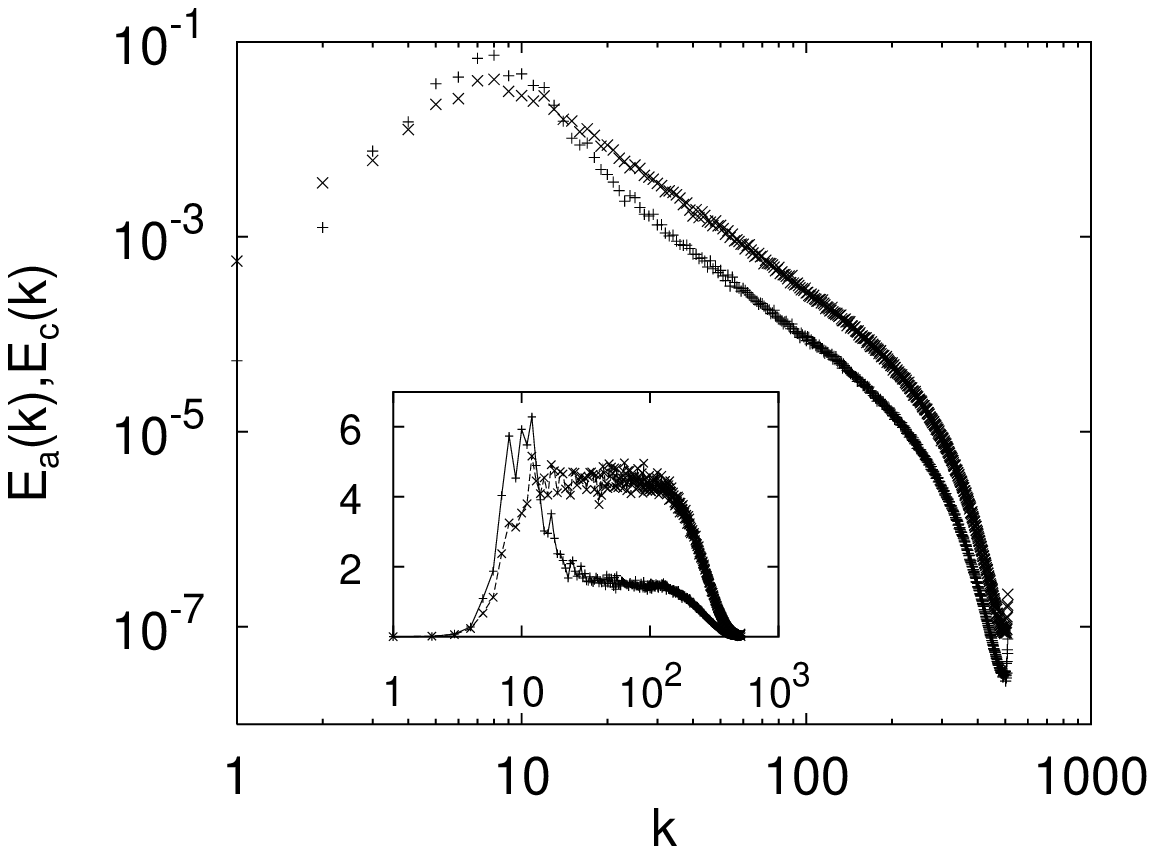} 
\caption{(a)The vorticity spectrum $E_a(k) \sim k^{-1-\xi}$ steepens
by increasing the Ekman coefficient $\alpha$. Here $\alpha = 0.15$
$(+)$, $\alpha = 0.23$ $(\times)$, $\alpha = 0.30$ $(\odot)$. In the
inset, the exponent $\xi$ as a function of $\alpha$. (b) Power
spectra of passive scalar ($\times$) and vorticity ($+$). Here
$\alpha=0.15$.  In the inset we show the same spectra compensated by
$k^{1+\zeta^{c}_2}$. For details on the DNS
see \protect\cite{BCMV02}.}
\label{fig:eck1} 
\end{figure}

\begin{figure}[t!]
\begin{center}
\includegraphics[draft=false, scale=0.74, clip=true]{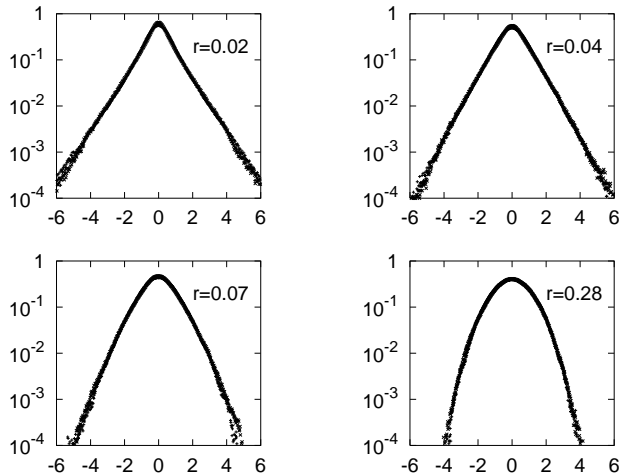} 
\caption{Probability density functions of vorticity differences
($+$) and of passive scalar ones ($\times$),
normalized by their respective standard deviation,
at different scales $r$ within the scaling range.}
\label{fig:eck3} 
\end{center}
\end{figure}

In the presence of friction ($\alpha>0$) kinetic energy
is removed at large scales holding the system in a statistically
steady state and small-scale statistics is
modified by the competition of the inertial (${\bm v}\cdot {\bm
\nabla} a$) and friction ($-\alpha \,a$).
Since these two terms have the same
dimension due to smoothness of the velocity field, this
results in nontrivial scaling laws for $a$. This effect
is evident in DNS with $\alpha>0$ \cite{NAGO00,BCMV02}, where the
vorticity spectrum displays a power law steeper than in the frictionless
case: $E_a(k)\sim k^{-1-\xi}$ with $\xi$ dependent on
$\alpha$ (see Fig.~\ref{fig:eck1}a).  For $0<\xi<2$ the exponent $\xi$
coincides with the scaling exponent, $\zeta^a_2$, of the second order
structure function $S^a_2(r)$. Additionally, high-order
structure functions at fixed $\alpha$ show anomalous scaling,
$\zeta^a_N\neq N\xi/2$ \cite{BCMV02}.
The spectral steepening and the presence of intermittency are observed
\cite{NAGO99,BCMV02} also in passive scalars evolving in smooth flows
according to
\begin{equation}
\partial_t c+{\bm v}\cdot{\bm \nabla}c=\kappa \Delta c-\alpha \,c+f_c\,
\label{eq:eck2}
\end{equation}
Physically Eq.~(\ref{eq:eck2}) describes the
evolution of a decaying passive substance (e.g. a radioactive marker)
\cite{MY75,C99}.  For $\alpha=0$ the dimensional expectation
 $E_c(k)\sim k^{-1}$ has been verified in
experiments \cite{JCT01}. For positive $\alpha$ passive scalar spectra
become steeper than $k^{-1}$ and, at high wavenumbers, have the same
slope of the vorticity spectrum (Fig.~\ref{fig:eck1}b). Additionally, the
pdfs of passive and vorticity increments for separations inside the
inertial range collapse one onto the others (Fig.~\ref{fig:eck3}),
signalling that the scaling exponents coincide: $\zeta^c_N=\zeta^a_N$. 
In summary, the vorticity
field and the passive scalar share the same statistical scaling properties
\cite{NAGO00,NAGO99,BCMV02}, similarly to the 2D convection case. However, 
differences may appear for odd order
moments \cite{B00}.

It is now interesting to understand how the equivalence of active and passive
statistics is realized: we shall see that in this case the smoothness
of the flow is a crucial ingredient.

Let us start with the decaying passive scalar (\ref{eq:eck2}).  First
of all, it should be noted that the presence of non-zero friction
($\alpha>0$) regularizes the field, and there is absence of
dissipative anomaly \cite{C99}, even if the mechanism is different
from the MHD one.  As a consequence, in Eq.~(\ref{eq:eck2}) we can put
$\kappa=0$ and solve it by the method of characteristics (see
Sect.~\ref{sec:lag}), i.e.  $c({\bm x},t)=\int_{-\infty}^t {\rm d}s
f_c({\bm \rho}(s;{\bm x},t),s) \exp(-\alpha(t-s))$, where now the path
${\bm \rho}(s;{\bm x},t),s)$ is unique due to the smoothness of the
velocity field (note that $E_v(k)\sim k^{-3-\xi}$ is always steeper
than $k^{-3}$, see also Fig.~\ref{fig:eck1}a). The integral extends up
to $-\infty$ where the initial conditions are set.

\begin{figure}[t!]
\begin{center}
\includegraphics[draft=false, scale=0.68, clip=true]{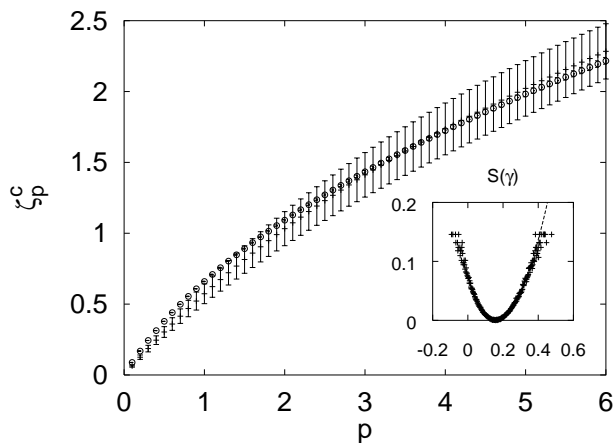} 
\caption{The scaling exponents of the passive scalar $\zeta^{c}_p$
($+$), which have been computed also for noninteger moments, indeed
Eq.~(\ref{eq:eck5}) holds in general.  We also show the exponents
obtained from the separation times statistics ($\odot$) according to
$\langle \exp[-\alpha p {\cal T}_{\ell_f}(r)] \rangle \sim
r^{\zeta_p^{c}}$ with average over about $2 \times 10^{5}$ couple
of Lagrangian particles.  The errorbars are estimated by the
r.m.s. fluctuation of the local slope. In the inset we plot the Cramer
function $S(\gamma)$ computed from finite time Lyapunov exponents
(symbols) and exit time statistics (line).}
\label{fig:eck4} 
\end{center}
\end{figure}

Passive scalar increments $\delta_r c= c({\bm x}_1,t)-c({\bm x}_2,t)$
($r=|{\bm x}_1-{\bm x}_2|$), which are the objects we are interested in,
are associated to particles pairs
\begin{equation}
\delta_r c =\!\!\!
\int_{-\infty}^t\!\!\! {\rm d}s e^{-\alpha(t-s)} [f_c({\bm
\rho}(s;{\bm x}_1,\!t),s)\!-\!\!f_c({\bm \rho}(s;{\bm x}_2,\!t),s)]\,.
\label{eq:eck3}
\end{equation} 
It should be now noted that the integrand $f_c({\bm \rho}(s;{\bm
x}_1,\!t),s)\!-\!\!f_c({\bm \rho}(s;{\bm x}_2,\!t),s)$ stays small as
long as the separation between the two paths remains below the forcing
correlation scale, $\ell_f$; while for separations larger than
$\ell_f$ it is approximately equal to a Gaussian random variable,
$X$. In the latter statement we used the independence between $f_c$
and the particle trajectories, ensured by the passive nature of
$c$. Therefore, Eq.~(\ref{eq:eck3}) can be approximated as
\begin{equation}
\delta_r c \approx X \int_{-\infty}^{t-{\cal T}_{\ell_f}(r)} \!\!
{\rm d}s \, e^{-\alpha (t-s)}  \sim X e^{-\alpha {\cal T}_{\ell_f}(r)}\,,
\label{eq:eck3bis}
\end{equation}
 where ${\cal T}_{\ell_f}(r)$ is the time necessary for the particles
pair to go from $r$ to $\ell_f$ backward in time. It is now clear that
large fluctuations are associated to fast separating couples, ${\cal
T}_{\ell_f}(r)\ll \langle{\cal T}_{\ell_f}(r)\rangle$, and small
fluctuations to slow ones. Moreover, since ${\bm v}$ is smooth,
two-dimensional and incompressible pairs separation is exponential
and its statistics is described by a single finite-time Lyapunov
exponent \cite{CPV91}, $\gamma$. It is related to the separation time 
through the relation:
\begin{equation}
\ell_f= r e^{\gamma{\cal T}_{\ell_f}(r)}\,.
\label{eq:eck4}
\end{equation}
 Large deviations theory states that at large times the random variable
$\gamma$ is distributed as $P(\gamma,t) \sim t^{1/2}
\exp[-S(\gamma)t]$.   $S(\gamma)$, the so-called Cramer function
\cite{E85}, is positive, concave and has a quadratic minimum,
$S(\lambda)=0$ in $\lambda$, the maximum Lyapunov
exponent. From (\ref{eq:eck3bis}) and  (\ref{eq:eck4}) along with
with the expression for $P(\gamma,t)$, structure functions can be
computed as
\begin{equation}
S^{c}_N(r) \sim \langle X^N \rangle 
\int {\rm d}\gamma \left({r\over
\ell_f}\right)^{{N\alpha+S(\gamma)\over \gamma}} \approx \left({r\over
\ell_f}\right)^{\zeta^c_N}\,,
\label{eq:eck5}
\end{equation}
where $\zeta^c_N=\min_{\gamma} \{N,[N\alpha+S(\gamma)]/\gamma\}$.
Fig.~\ref{fig:eck4} shows that this prediction is verified by
numerical simulations. 
The anomalous exponents $\zeta^c_N$ can therefore be evaluated 
in terms of the flow properties through the Cram\'er funtion $S(\gamma)$. 

Since $\zeta^c_N=\zeta^a_N$ one may be tempted to apply straightforwardly 
to vorticity the
same argument used for the passive scalar.  However,
the crucial assumption used in the derivation of Eq.~(\ref{eq:eck5}) is the
statistical independence of particle trajectories and the forcing.
That allows to consider ${\cal T}_{\ell_f}(r)$ and $X$ as independent
random variables.  While this is obviously true for $c$, it is clear
that the vorticity forcing, $f_a$, may influence the velocity statistics
and, as a consequence, the particle paths. In other terms, for the
vorticity field we cannot {\it a priori} consider $X$ and ${\cal
T}_{\ell_f}(r)$ as independent.

Nevertheless, it is possible to argue in favour of the validity of 
Eq.~(\ref{eq:eck3}) also
for vorticity. Indeed, the random variable ${\cal
T}_{\ell_f}(r)$ is essentially determined by the evolution of the
strain along the Lagrangian paths for times $t-{\cal
T}_{\ell_f}(r)<s<t$, whilst $X$ results from the scalar input accumulated at
times $s<t-{\cal T}_{\ell_f}(r)$.  Since the strain correlation time
is basically $\alpha^{-1}$, for ${\cal T}_{\ell_f}(r)\gg \alpha^{-1}$
it is reasonable to assume that ${\cal T}_{\ell_f}(r)$ and $X$ are
statistically independent.  Eq.~(\ref{eq:eck4}) allows to translate 
the condition ${\cal T}_{\ell_f}(r)\gg \alpha^{-1}$
into $r\ll \ell_f \exp(-\gamma/\alpha)$, and to conlcude that at small scales
we can safely assume that the vorticity behaves as a passive field.

We conclude this section with two remarks.  First, a difference
between the present scenario and the one arisen in two-dimensional
convection should be pointed out.  Here the scaling exponents depend
on the statistics of the finite time Lyapunov exponents, which in turn
depends on the way the vorticity is forced.  As a consequence
universality may be lost.  Second, the smoothness of the velocity
field plays a central role in the argument used to justify the
equivalence of the statistics of $a$ and $c$ in this system. For rough
velocity fields ${\cal T}_{\ell_f}(r)$ is basically independent of $r$
for $r << \ell_f$ and the argument given in this section would not
apply.

\section{Turbulence on fluid surfaces}
\label{sec:5}
\subsection{Surface Quasi Geostrophic turbulence}
The study of geophysical fluids is of obvious importance for the
understanding of weather dynamics and global circulation.  Taking
advantage of stratification and rotation, controlled approximations on
the Navier-Stokes equations are usually done to obtain more tractable
models.  For example, the motion of large portions of the atmosphere
and ocean have a stable density stratification, i.e. lighter fluid
sits above heavier one. This stable stratification, combined with the
planetary rotation and the consequent Coriolis force, causes the most
energetic motion to occur approximately in horizontal planes and leads
to a balance of pressure gradients and inertial forces. This situation
is mathematically described by the Quasi-Geostrophic equations
\cite{S98,R79}.  On the lower boundary, the surface of the earth or
the bottom of the ocean, the vertical velocity has to be zero and this
further simplifies the equations. Assuming that the fluid is
infinitely high and that the free surface is flat, the dynamics is
then described by the Surface Quasi-Geostrophic equation (SQG)
\cite{PHS94,C02}. It governs the temporal variations of the density
field, our active scalar $a$ throughout this section.  For an ideal
fluid the density variation will be proportional to the temperature
variation and one can use the potential temperature as fundamental
field.

The density fluctuations evolve according to the transport equation
 \begin{equation} 
\partial_t a+{\bm v} \cdot {\bm \nabla} a = \kappa \Delta a+f_a\,,
\label{eq:sqg1}
\end{equation}
and the velocity field is functionally related to $a$.  In terms of
the stream function $\psi$, the density $a$ is obtained as $a({\bm
x},t)=(-\Delta)^{1/2} \psi({\bm x},t)$, and ${\bm v}=
(\partial_y,-\partial_x)\psi$.  In Fourier space the link is
particularly simple
\begin{equation}
\hat {\bm v}({\bm k},t)=-i \left({k_y\over k},-{k_x\over k}\right) \hat
a({\bm k},t)\,.
\label{eq:sqg2}
\end{equation}
In the early days of computer simulations eqs.~(\ref{eq:sqg1}) and
(\ref{eq:sqg2}) had been used for weather forecasting.  Recently they
received renewed interest in view of their formal analogy with the
$3d$ Navier-Stokes equations \cite{C95,MT96}, and as a model of active
scalar \cite{PHS94,C98}

It is instructive to generalize  (\ref{eq:sqg2}) as 
\begin{equation}
\hat {\bm v}({\bm k},t)=-i \left({k_y\over k^z},-{k_x\over k^z}\right)
\hat a({\bm k},t)\,.
\label{eq:sqg3}
\end{equation}
For $z=1$ one recovers (\ref{eq:sqg2}), while $z=2$ corresponds to the
$2d$ Navier-Stokes equation and $a$ is the vorticity (see the previous
section).  The value of $z$ tunes the degree of {\it locality}.  The
case of interest here is  $z=1$, corresponding to the same degree of
locality as in $3d$ turbulence \cite{PHS94}.

Eqs.(\ref{eq:sqg1}),(\ref{eq:sqg3}) have two inviscid unforced
quadratic invariants $E=-\int {\rm d}{\bm x} a\psi$ and $\Omega=\int
{\rm d}{\bm x} a^2$, which for $z=2$ correspond to the energy and the
enstrophy, respectively.  Notice that, for $z=1$, the fields ${\bm v}$
and $a$ have the same dimension and $E_a(k)\equiv
E_v(k)$. Quasi-equilibrium arguments predict an inverse cascade of $E$
and a direct cascade of $\Omega$. If $k_f$ identifies the injection
wave number dimensional arguments give the expected spectral behavior
\cite{PHS94}:
\begin{equation}
E_a(k)\propto\left\{
\begin{array}{cc}
k^{-{7 \over 3}+{4z\over 3}} \qquad k\ll k_f\\
k^{-{7 \over 3}+{2z\over 3}} \qquad k\gg k_f
\end{array}
\right.\,.
\label{eq:sqg4}
\end{equation}
For $z=2$ one recovers the $2d$ Navier-Stokes expectation
\cite{kraichnan67,batchelor69}. Here, we are interested in the range
$k>k_f$ and $z=1$, so that $E_a(k)\equiv E_v(k)\sim k^{-5/3}$ as in
$3d$ turbulence.  Assuming absence of intermittency, the spectrum
$E_v(k)\sim k^{-5/3}$ would imply $\delta_r v\sim r^{1/3}$.
Therefore, for a passive scalar one expects the standard
phenomenology: an intermittent direct cascade with $E_c(k)\sim
k^{-5/3}$ to the Oboukov-Corrsin-Kolmogorov type of arguments
\cite{MY75}.

\subsubsection{Direct numerical simulations settings}

We performed a set of high resolution direct numerical simulations
of Eqs.~(\ref{eq:sqg1}) and (\ref{eq:sqg2}) along with the passive
scalar equation (\ref{eq:2}). Integration has been performed by means
of a $2/3$ de-aliased pseudo-spectral method on a $2\pi\times 2\pi$
doubly periodic square domain with ${\cal N}^2$ collocation
points. Time integration has been done with a second-order
Adam-Bashforth or Runge-Kutta algorithms, approriately modified to exactly
integrate the dissipative terms. The latter ones, as customary,
have been replaced by a hyper-diffusive term that in Fourier space
reads $- k^{2p} \hat{a}({\bm k},t)$. Since the system is
not stationary due to the inverse cascade of $E$, we added an energy
sink at large scales in
 the form $-k^{-q}\hat{a}({\bm k},t)$. In order to
evaluate its possible effect on inertial quantities, a very high
resolution ($4096^2$) DNS has been done without any energy sink
at large scales.

\begin{table}
\caption{\label{tab:1} Summary of settings in DNS.} 

\begin{indented}
\lineup
\item[]\begin{tabular}{@{}*{6}{l}}
\br                              
$\0\0run$ &${\cal N}^2$ &$p$  &$q$ &$F$  &$k_f$\cr 
\mr
$\0\01^a$ & $4096^2$ &2 &$\;$    &B    &44-48 \cr
$\0\02^b$ & $2048^2$ &2 &1/2  &B    &2-6   \cr
$\0\03^b$ & $2048^2$ &2 &0    &G    &5     \cr
$\0\04^b$ & $2048^2$ &2 &1/2  &G    &5     \cr
$\0\05^c$ & $2048^2$ &2 &0    &NG   &5     \cr
\br
\end{tabular}
\item[] $^{\rm a}$ Runs 1,2 are forced according to
Eq.~(\ref{eq:forcings}), and run 1 is without any friction term. 
\item[] $^{\rm b}$ Run 3,4 are forced according to Eq.~(\ref{eq:fband}).
\item[] $^{\rm c}$ Run
5 is forced with the non-Gaussian forcing discussed in
Sect.~\ref{sec:3points}. Lower ($512^2$) resolutions runs with
several settings both for the dissipative and friction terms have been
also performed
\end{indented}
\end{table}

A summary of the numerical settings can be found in
table~\ref{tab:1}. We considered different scalar inputs: (G) a
Gaussian $\delta$-correlated in time forcing as (\ref{eq:forcings});
(B) a $\delta$-correlated in time one restricted to a few wavenumbers
shells as (\ref{eq:fband}); (NG) a non-Gaussian non
$\delta$-correlated in time one suited to produce non zero three
points correlations for the scalar fields (see below in
Sect.~\ref{sec:3points}).
\subsection{Active and passive scalar statistics in SQG turbulence}
\begin{figure}[t!]
\includegraphics[draft=false, scale=0.63, clip=true]{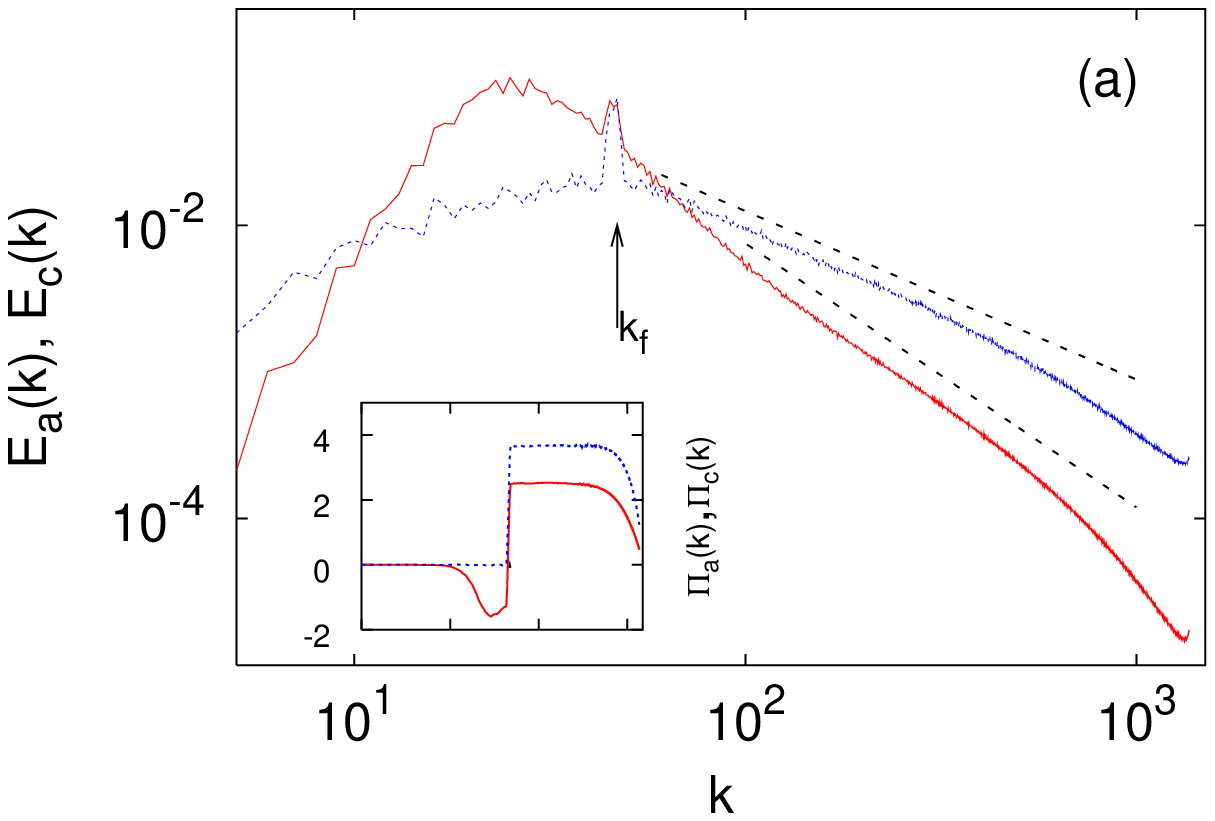} 
\includegraphics[draft=false, scale=0.63, clip=true]{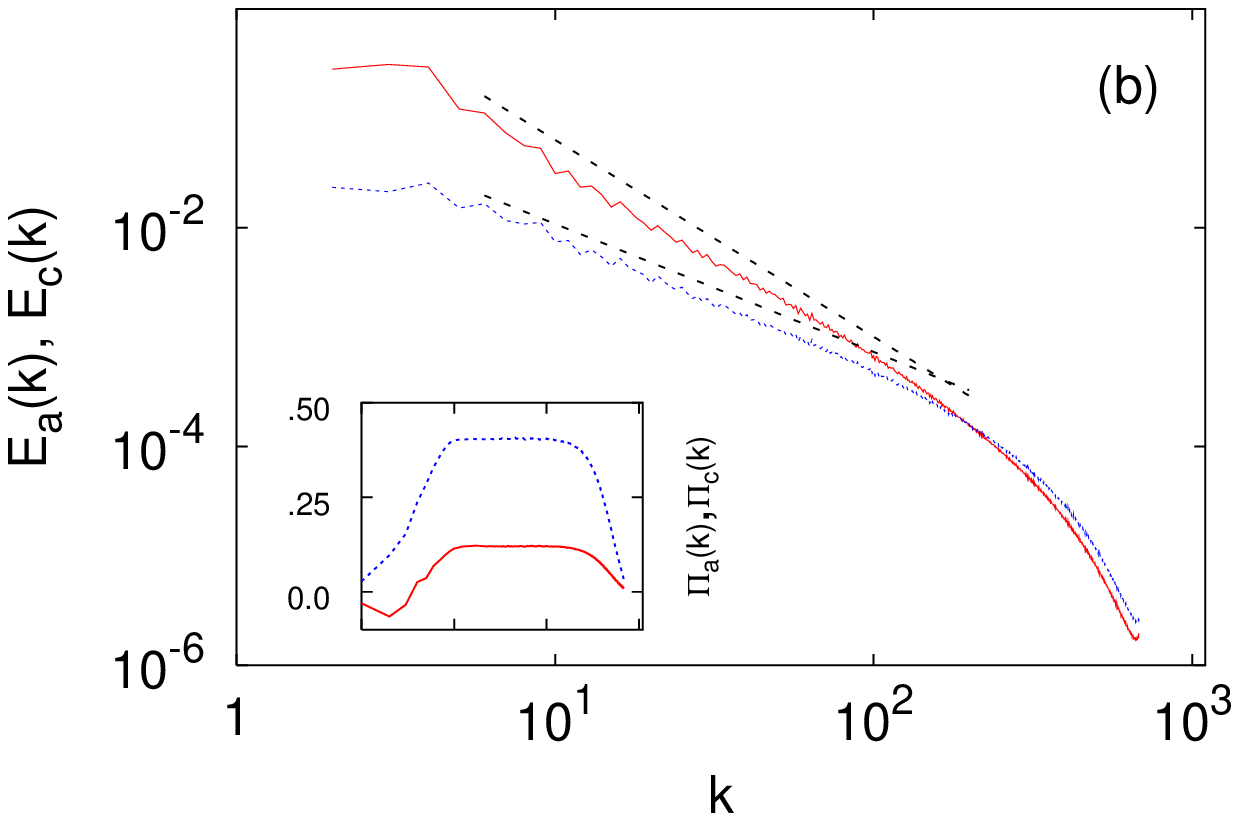} 
\caption{(a) Power spectra of active (red) and passive (blue) scalar
variances $E_a(k)=\pi k |\hat{a}({\bm k},t)|^2$ and $E_c(k)=\pi k
|\hat{c}({\bm k},t)|^2$ for run~1, the inset shows the active, and
passive energy fluxes $\Pi_{a,c}(k)$. The dashed lines corresponds to
the best fitted spectral slopes $E_{a}(k) \sim k^{-1.8\pm 0.1}$ and
$E_{c}(k) \sim k^{-1.15\pm 0.05}$. Note that the fitted slope for
$E_a(k)$ is better recovered at higher wavenumbers, close to the
energy peak a steeper ($\sim k^{-2\pm 0.1}$) is observed.  (b) The
same but for run~3. The left inset shows the scalar fluxes.  The
dashed lines indicates the spectral slopes $E_a(k)\sim k^{-1.8 \pm
0.1}$ and $E_c(k)\sim k^{-1.17\pm 0.05}$. Here the scaling range is
wider than in run~1. For both runs active and passive spectra have
been shifted for visualization purposes. In the other runs we observe
the same qualitative and quantitative features, within the error
bars. In particular, run~1 and run~2 give almost indistinguishable
spectral slopes, meaning that the large scale energy sink does not
enter too much into the inertial range, as expected since the velocity
field is rough at small scales.  Run~5 produced spectral slopes
indistinguishable from those of runs~3 and 4.  However, it should be
noted that run~1 and run~2 display a steeper spectra close to the forcing wave
numbers, this feature is absent in the other runs and may be 
due to finite size effects, see text and
also Fig.~\ref{fig:phb3}.  }
\label{fig:phb1}
\end{figure}

Let us start looking at the measured scalar spectra in order to check
the dimensional predictions. In Fig.~\ref{fig:phb1} we summarize the
spectral behavior for $a$ and $c$ in two different simulations
(i.e. runs 1 and 3).  

Two observations are in order:

 {$\bullet$} The scalar spectra $E_a(k)$ is steeper
than the dimensionally expected $k^{-5/3}$. The slope
does not seem to depend on the injection mechanism. 

{$\bullet$} The active
and passive scalars, both performing a direct cascade, are different
already at the level of the spectral slopes. In particular, $c$ is
much rougher than $a$.

\begin{figure}[t!]
\begin{center}
\includegraphics[draft=false, scale=0.48, clip=true]{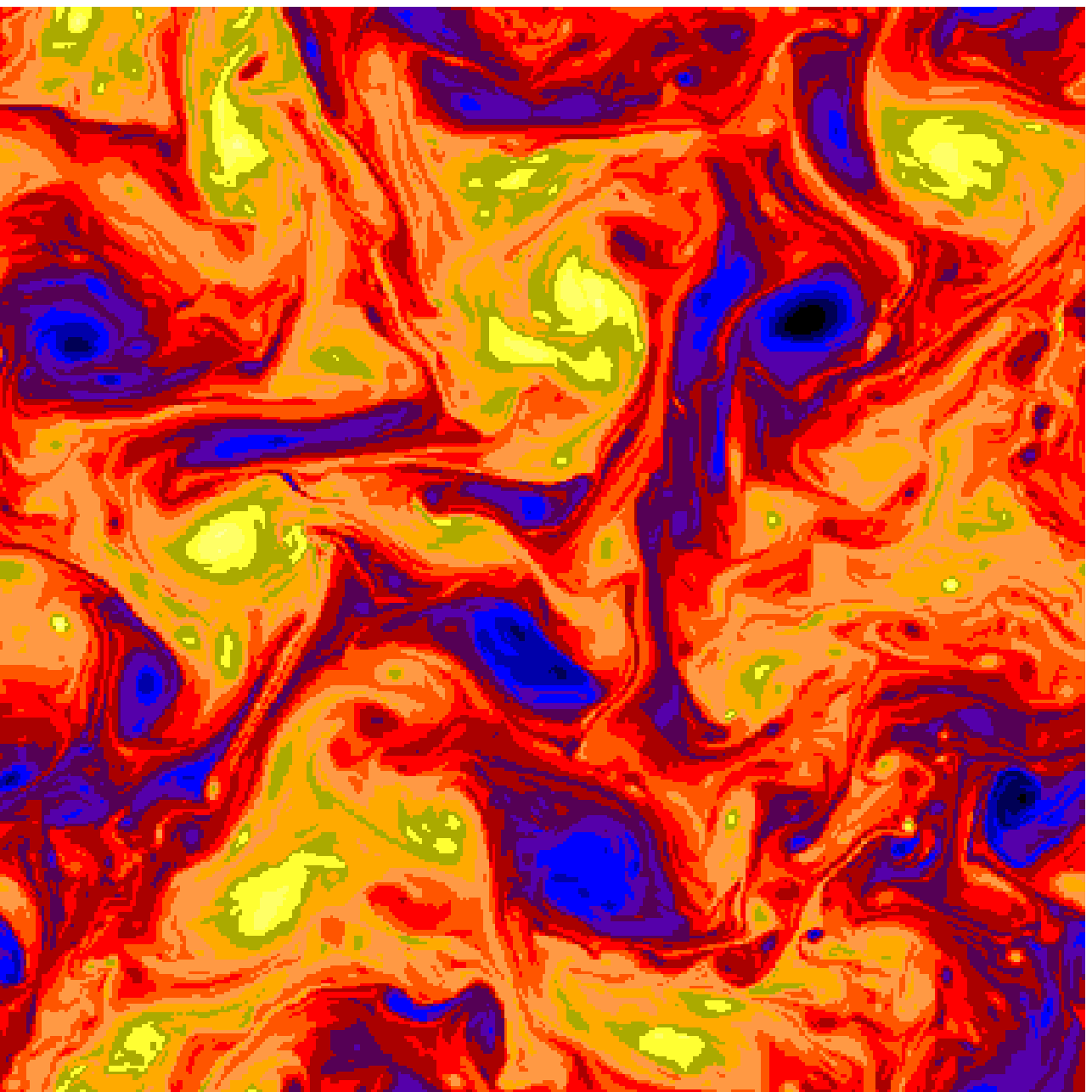} 
\hspace{.5cm}\includegraphics[draft=false, scale=0.48, clip=true]{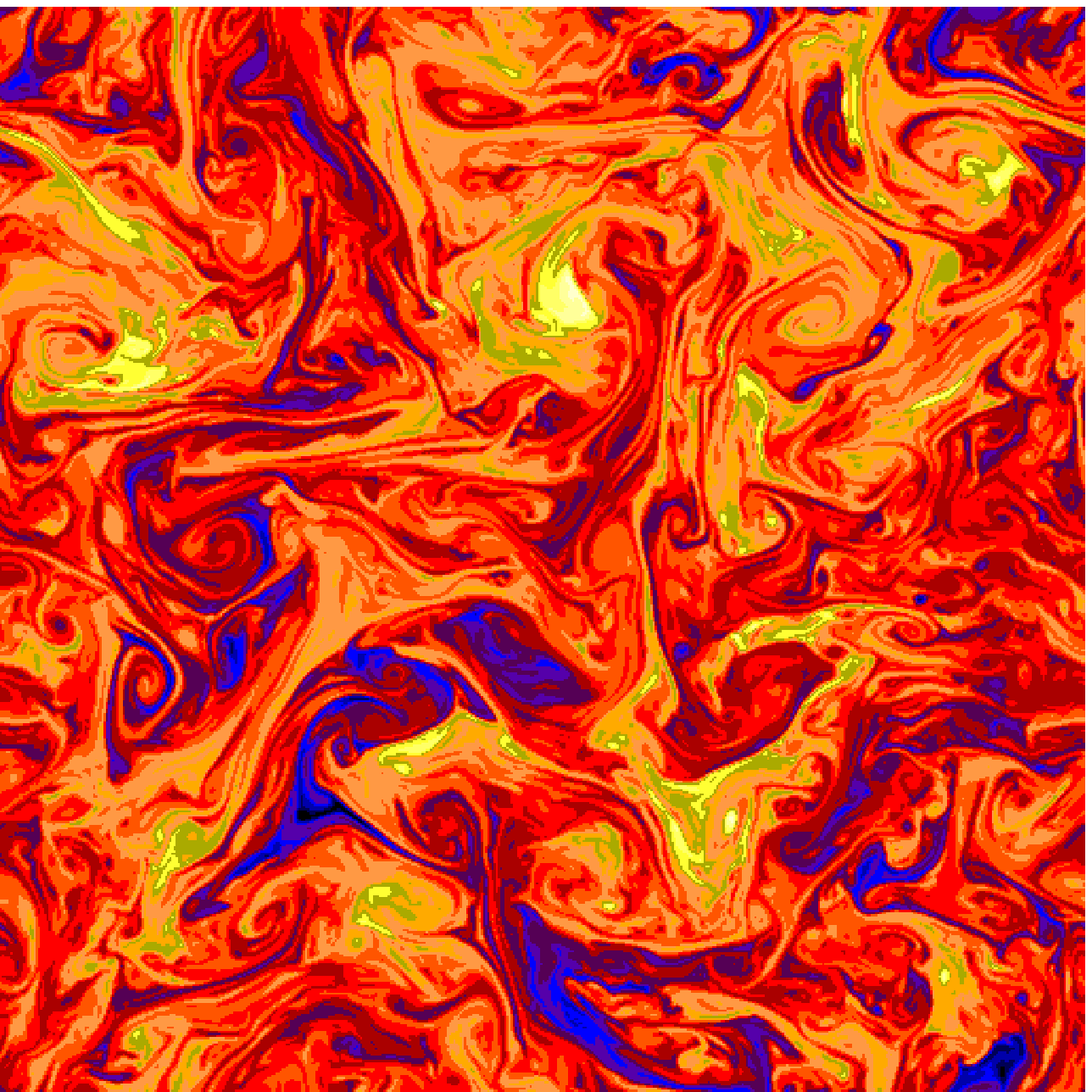} 
\caption{Snapshots of the active (left) and passive (right) scalar (at
$512^2$ resolution). Note the presence of coherent structures in the
active field, which are slow evolving. The passive field displays
filamental like features indicating that it is more rough than the active
one.}
\label{fig:phb2} 
\end{center}
\end{figure}

The differences between $a$ and $c$ are evident also from
Fig.~\ref{fig:phb2}, where two simultaneous snapshots of the fields
are displayed. A direct inspection of the fields confirms that $c$ is
more rough than $a$, and resembles a passive scalar in a smooth
flow. Moreover, in $a$ we observe the presence of large coherent
structures at the scale of the forcing.  These are actually
long-lived, slowly-evolving structures that strongly impair the
convergence of the statistics for high-order structure functions. In
the following we thus limit ourselves to a comparison of the low-order
statistics of $a$ and $c$. That is however sufficient to appreciate
the differences beween active and passive scalar statistics.

\begin{figure}[b!]
\includegraphics[draft=false, scale=0.63, clip=true]{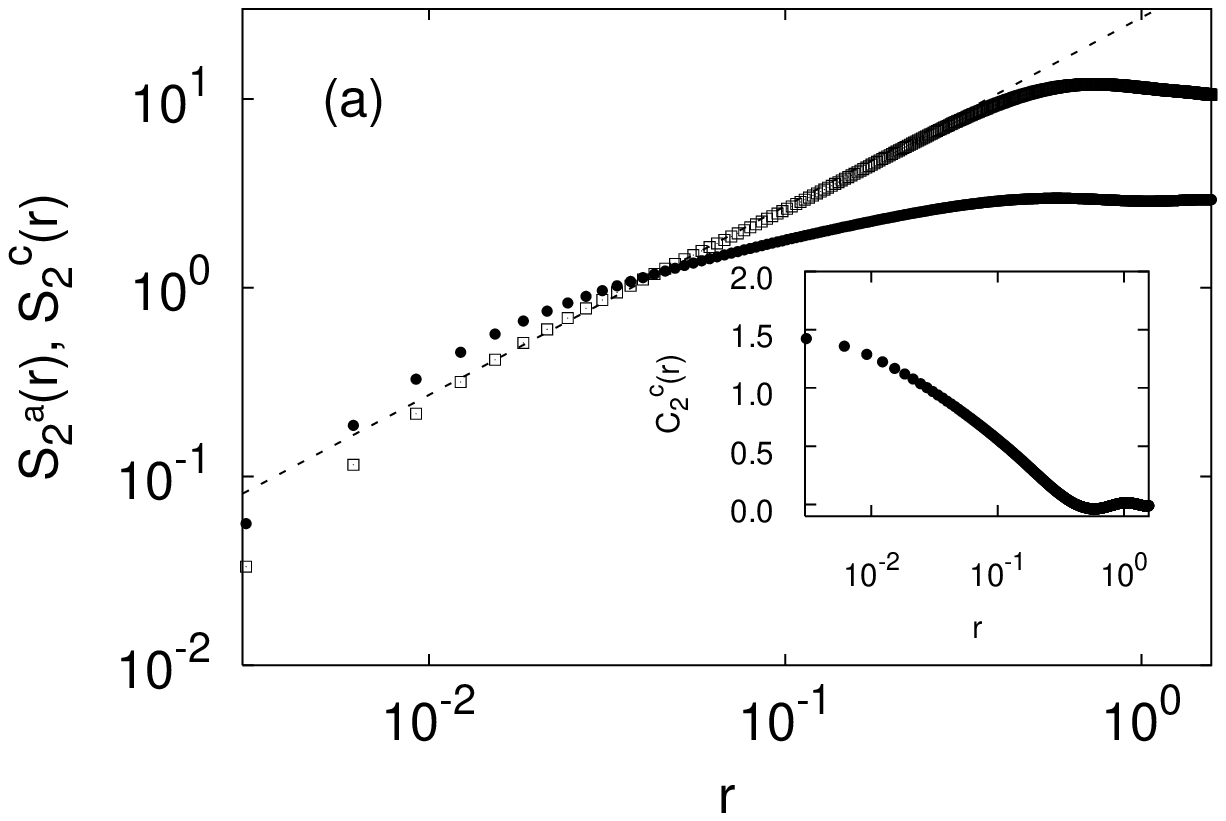} 
\includegraphics[draft=false, scale=0.63, clip=true]{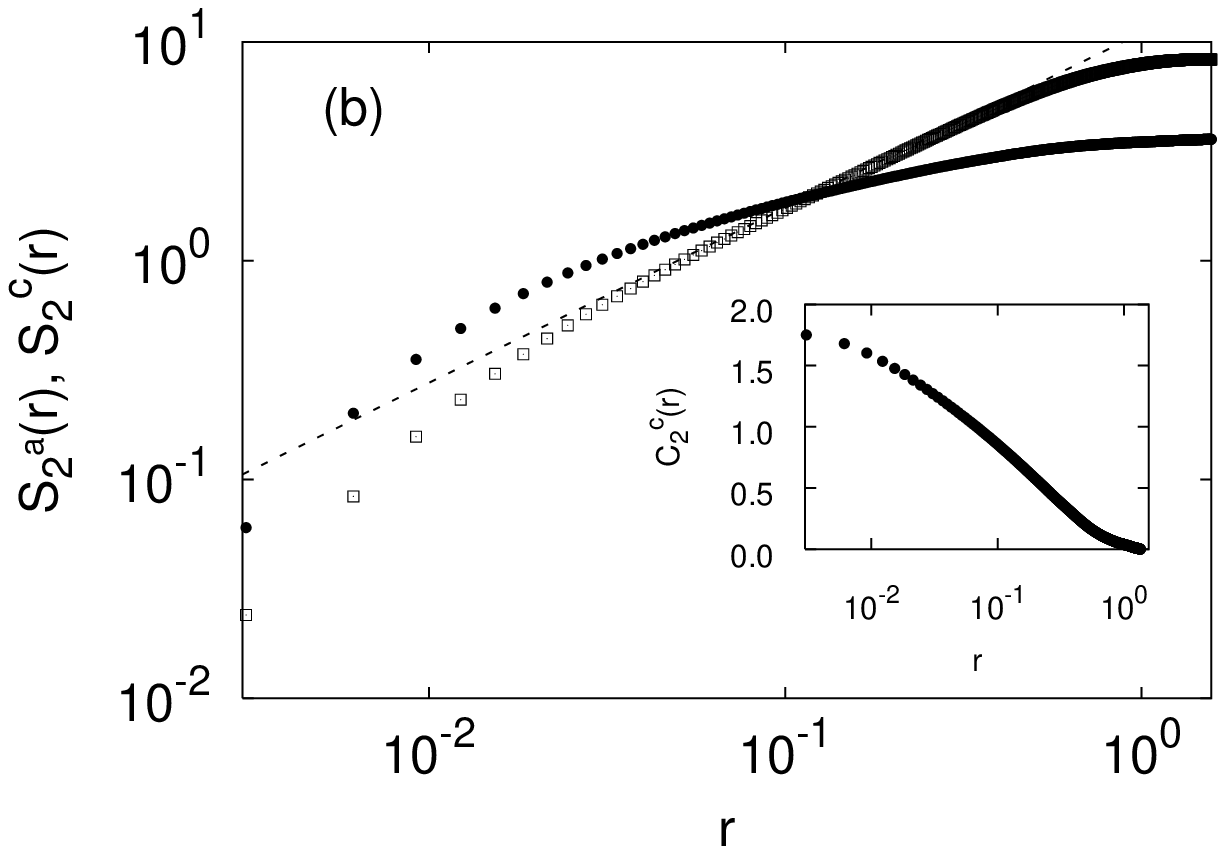} 
\caption{ (a) Second order structure function for the active (empty
boxes) and the passive (full circles) scalar in run 2. The straight
line shows the scaling $\sim r^{1}$ for the $S^a_2(r)$ (the same
exponent is observed in run 1).  The passive scalar does not display
a neat scaling behavior.  In the inset we show the correlation
function for the passive scalar, ${\cal C}^c_2(r)$, in log-lin scale,
(see text). (b) The same but for run 3.  Here the straight dashed line
indicates the slope $r^{0.8}$.  $S^a_2(r)$ in run 5 has a slope $\sim
r^{0.84}$, compatible with the ones
observed in runs 3 and 4 within statistical errors.}  
\label{fig:phb3} 
\end{figure}

The deviation from the dimensional expectation for the spectral slope
of the active field was already observed in previous numerical studies
\cite{PHS94}, and it is possibly due to intermittency in the active
scalar and velocity statistics. Indeed, the rescaled pdfs of the
increments do not collapse (not shown).  Concerning the universality
of anomalous exponents with respect to the forcing statistics, we
observe that the spectral slopes do not depend sensitively on the
forcing statistics.  However, the forcing (B) (see table~\ref{tab:1})
induces a steeper spectrum close to the forcing scale, while a
universal slope is apparently recovered at large wavenumbers.
(Fig.~\ref{fig:phb1}a).  The scaling of $S^a_2(r)$ seems to be more
sensitive to the forcing (see Fig.~\ref{fig:phb3}). These
discrepancies may be due to finite size effects which are more severe
for forcing (B) which is characterized by a slower decay of the
spatial correlations.  Unfortunately, it is difficult to extract
reliable information on the high-order statistics. Indeed the presence
of slowly evolving coherent structures induces a poorly converging
statistics for high-order structure functions.  Therefore, we cannot
rule out the possibility of a dependence of active scalar exponents on
forcing statistics.

\begin{figure}[t!]
\begin{center}
\includegraphics[draft=false, scale=0.48, clip=true]{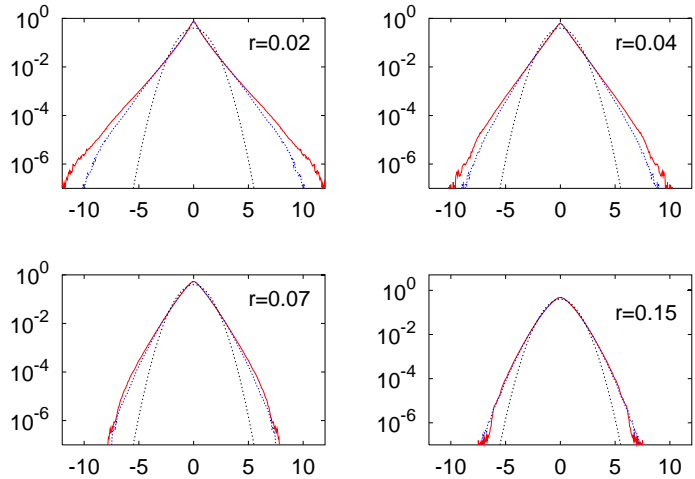} 
\caption{Pdf of active (red) and passive (blue) scalar differences
normalized by their respective standard deviations for four different
scales within the inertial range. The dotted lines are Gaussian
pdfs drawn for comparison. Data refer to run 3, the other runs 
show similar results.} 
\label{fig:phb4} 
\end{center}
\end{figure}

A result that emerges beyond any doubt is that active and passive
scalars behave differently, as shown in Figs.~\ref{fig:phb1},
\ref{fig:phb2} and \ref{fig:phb3}.  This is confirmed by the
differences in the pdfs of $\delta_r a$ and $\delta_r c$ at various
scales within the inertial range (Fig.~\ref{fig:phb4}). The
single-point pdfs of $a$ and $c$ are different as well (not shown).

It is worth noticing that the behavior of the passive scalar deviates
from na\"{\i}ve expectations. We observe $E_c(k)\sim k^{-1.15}$
whereas a dimensional argument based on the observed velocity spectrum
($E_v(k)\equiv E_a(k)\sim k^{-1.8}$) and on the Yaglom relation
\cite{MY75} ($\langle \delta_r v \,(\delta_r c)^2\rangle\sim r$) would
give $E_a(k)\sim k^{-1.6}$.  The violation of the dimensional
prediction is even more striking looking at $S^c_2(r)$ in
Fig.~\ref{fig:phb3}.  This feature is reminiscent of some experimental
investigations of passive scalars in turbulent flows, see e.g.
Refs.~\cite{VID01,VI99} and references therein, where shallow scalar
spectra are observed for the scalar even at those scales where the
velocity field displays a K41 spectrum.  It is likely that the
presence of coherent structures (see Fig.~\ref{fig:phb2}) leads to
persistent regions where the velocity field has a smooth (shear-like)
behavior. This suggest a two-fluid picture: a slowly evolving
shear-like flow, with superimposed faster turbulent
fluctuations. Under those conditions, one may expect that the particle
pairs separate faster than expected, leading to shallower passive
scalar spectra \cite{SHEAR}.

In conclusion, even though both passive and active scalar perform a
direct cascade, their statistical properties are definitely different.

\subsection{Scaling and geometry}
\label{sec:3points}
Let us now compare the two scalars field by investigating
the three-point correlation functions of the active field, ${\cal
C}^{a}_3({\bm x}_1,{\bm x}_2,{\bm x}_3)=\langle a({\bm x}_1,t) a({\bm
x}_2,t)a({\bm x}_3,t)\rangle$, and of the passive one, ${\cal
C}^{c}_3({\bm x}_1,{\bm x}_2,{\bm x}_3)=\langle c({\bm x}_1,t) c({\bm
x}_3,t)c({\bm x}_3,t)\rangle$. This allows to compare the
scaling properties of the
correlation function by measuring the dependence 
on the global size of the triangle
identified by the three points, $R^2=\sum_{i<j} x_{ij}^2$ (being
$x_{ij}=|{\bm x}_i-{\bm x}_j|$). We will also investigate
 the geometrical dependence of $C_3$.

It has to be noted that for $\delta$-correlated Gaussian forcings as
(\ref{eq:forcings}) and (\ref{eq:fband}), ${\cal C}_3^{a}$ and ${\cal
C}_3^{c}$ are identically zero. Therefore we need a different forcing
statistics in order to study three-point correlations.  A possibility
is to break the rotational symmetry of the system by an anisotropic
forcing, e.g., a mean gradient (\ref{eq:mg}) as in
\cite{CV01}. However, that choice leads inevitably to $a=c$: the
equations are identical for $a$ and $c$ so that the difference field
$a-c$ will decay out.  We then choose to force the system as
follows. The two inputs $f_a$ and $f_c$ are
\begin{equation}
f_{a,c}({\bm x},t)= g_{a,c}^2({\bm x},t)-\int {\rm d} {\bm y}
g_{a,c}^2({\bm y},t)
\end{equation} 
where $g_{a,c}$ is a homogeneous, isotropic and 
Gaussian random field with correlation
\begin{equation}
\langle g_i({\bm x},t) g_j({\bm x}',t')\rangle=
\delta_{i,j}{\cal G}(|{\bm x}-{\bm x}'|/\ell_f) e^{-{|t-t'| \over \tau_f}}\,,
\end{equation}
where $i,j=a,c$, $\ell_f$ is the forcing scale, $\tau_f$ the forcing
correlation time and ${\cal G}(r)\propto G_0 \exp(-r^2/2)$.  The time
correlation is imposed by performing an independent Ornstein-Uhlenbeck
process at each Fourier mode, i.e. integrating the stochastic
differential equation $d\hat{g_{i}}({\bm k},t)=-1/\tau_f
\hat{g_{i}}({\bm k},t)dt +\sqrt{2G_0 dt/\tau_f} dw_i({\bm k},t)$
(where $dw_i$ are zero mean Gaussian variables with $\langle dw_i({\bm
k},t)dw_j({\bm k}',t')\rangle=\delta_{i,j} \delta(t-t')\delta({\bm
k}-{\bm k}')$). 
If $\tau_f \ll{\cal T}_{\ell_f} $ by the central
limit theorem a Gaussian statistics is recovered. Therefore, we
fixed $\tau_f\sim O({\cal T}_{\ell_f})$ in our DNS.
The advantages of this choice are that it preserves the isotropy and
gives analytical control on the forcing correlation functions.

\begin{figure}[t!]
\includegraphics[draft=false, scale=0.25, clip=true]{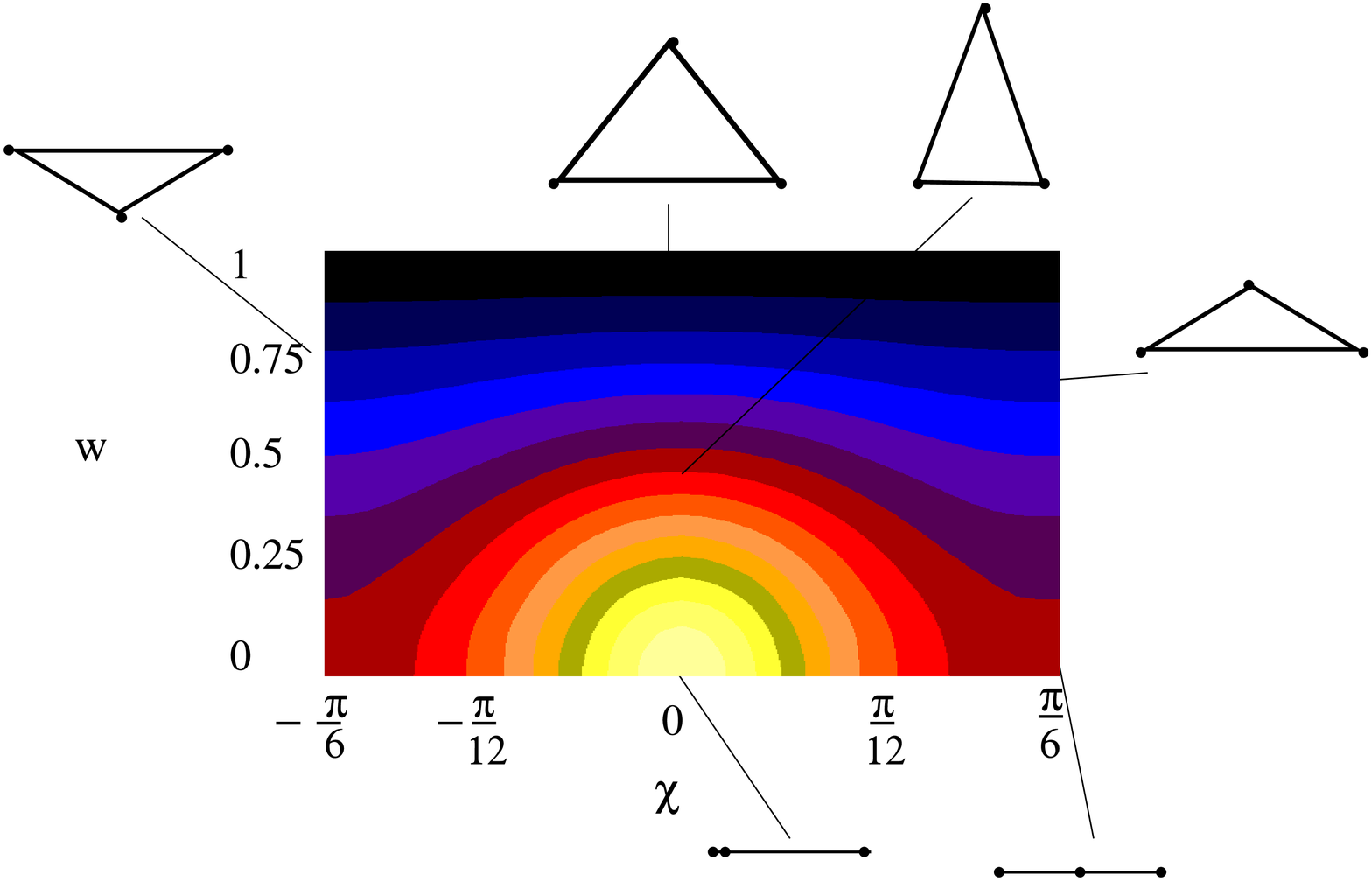}
\includegraphics[draft=false, scale=0.25, clip=true]{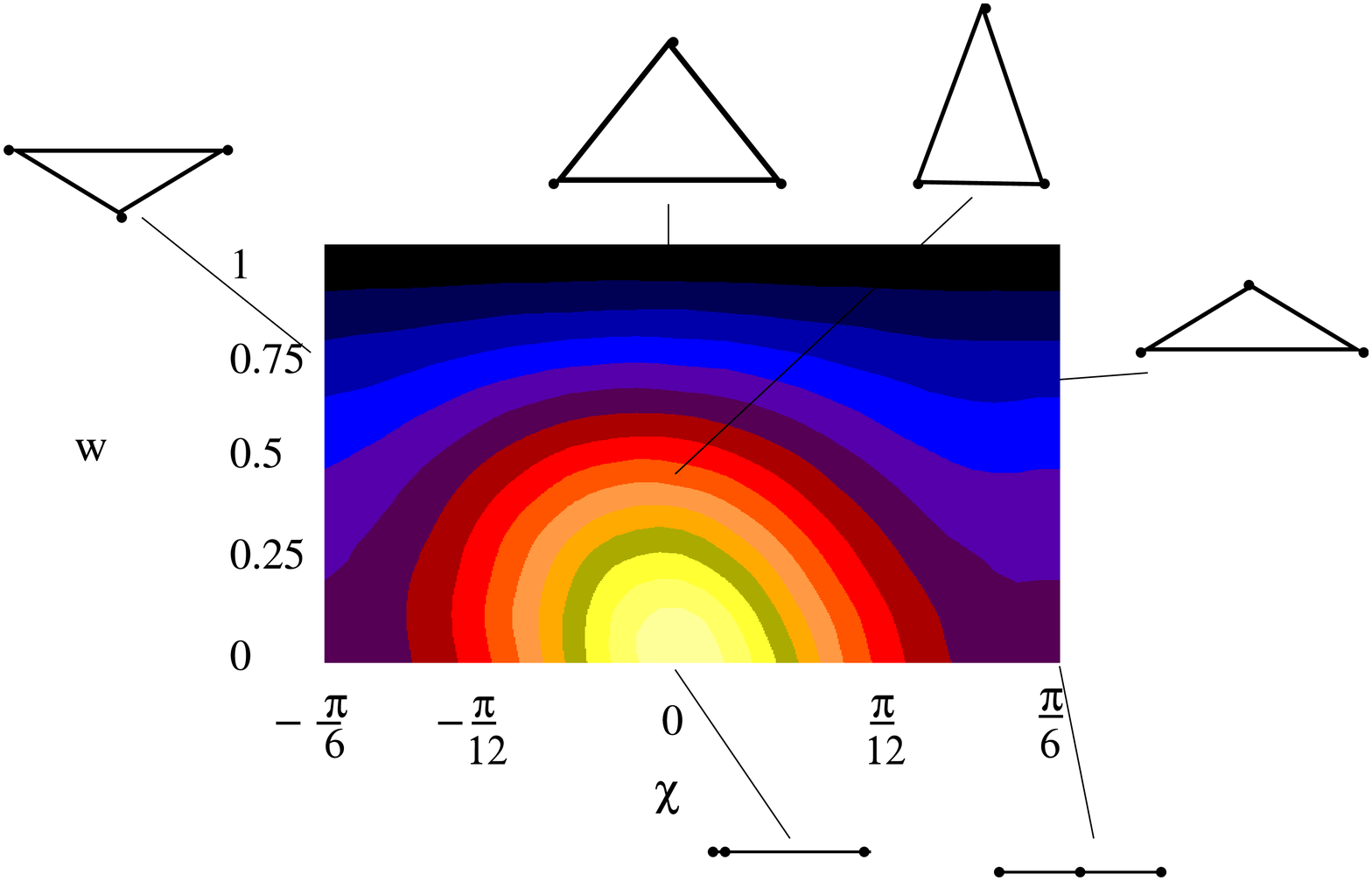} 
\caption{\label{fig:phb5} Contour plots in the $\chi-w$ plane of the
three-point shape function, $\Phi$, for passive (left) and active
(right) scalar.}
\end{figure}

Let us now see how the triangle identified by the three points,
$\underline{\bm x}=({\bm x}_1,{\bm x}_2,{\bm x}_3)$ can be
parameterized.  In two dimensions we need $3d=6$ variables to define a
triangle.  Since the correlation functions should possess all the
statistical symmetries of the system the number of degrees of freedom
is reduced. In particular, translational invariance ensures no
dependence on the position of the center of mass of the triangle,
$({\bm x}_1+{\bm x}_2+{\bm x}_3)/3$. The correlation function is thus
a function of the separation vectors among the $3$ points, i.e.
${\cal C}_3^{a,c}(\underline{\bm x})={\cal C}_3^{a,c}({\bm
x}_{12},{\bm x}_{23},{\bm x}_{31})$. Additionally, isotropy implies
that a rigid rotation of the triangle has no effect on the value of
$C_3$, so that three variables suffice: the global size of the
triangle $R$, and two parameters that define its shape.  In terms of
the Euler parametrization~\cite{SS98,P98,CP01}, upon defining ${\bm
\rho}_1=({\bm x}_1-{\bm x}_2)/\sqrt{2}$ and ${\bm \rho}_2=({\bm
x}_1+{\bm x}_2-2{\bm x}_3)/\sqrt{6}$, the shape of the triangle is
given in terms of the two variables:
\begin{equation}
w={2 {\bm \rho}_1\times{\bm \rho}_2\over R^2}\, \qquad 
\chi={1\over 2} \tan^{-1}\left[{2\,{\bm \rho}_1 \cdot {\bm \rho}_2 \over \rho_1^2-\rho_2^2}\right]\,,
\end{equation}
where $|w|$ is the ratio of the area of the triangle divided by the
area of the equilateral triangle having the same size,
$R^2=\rho_1^2+\rho_2^2$. $\chi$ has not a simple geometrical
interpretation.  Some shapes corresponding to a few $(w,\chi)$
are shown in Fig.~\ref{fig:phb5}.

The three-point correlation function can be 
decomposed as \cite{BGK98,FGV01}
\begin{equation}
{\cal C}^{a,c}_3(\underline{\bm x})= {\cal R}\{{\cal C}^{a,c}_3
(\underline{\bm x})\}+{\cal I}\{{\cal C}^{a,c}_3
(\underline{\bm x})\}\,,
\end{equation}
where ${\cal R},{\cal I}$ are the reducible and
irreducible components, respectively.  This means that ${\cal C}_3$
can be expressed as the sum of a part that depends on three points
(hereafter denoted as $C_3$) and a part that depends on two points
(hereafter denoted $C_2$), i.e. ${\cal C}_3({\bm
x}_1,{\bm x}_2,{\bm x}_3)= C_3({\bm x}_1,{\bm x}_2,{\bm x}_3)+C_2({\bm
x}_1,{\bm x}_2)+C_2({\bm x}_2,{\bm x}_3)+C_2({\bm x}_3,{\bm
x}_1)$. 
The reducible, ${\cal R}\{{\cal C}_3\}=
C_2({\bm x}_1,{\bm x}_2)+C_2({\bm x}_2,{\bm x}_3)+C_2({\bm x}_3,{\bm
x}_1)$ and irriducible part ${\cal I}\{{\cal C}_3\}=C_3({\bm x}_1,{\bm
x}_2,{\bm x}_3)$ are in general characterized by different scaling 
properties as a function of the triangle size $R$, and different
geometrical (shape) dependencies \cite{GLP97,CV01}. 

In terms of the variables $R,w$ and $\chi$, the reducible and
irreducible components take the following form
\begin{eqnarray}
{\cal R}\{{\cal C}^{a,c}_3(\underline{\bm x})\}&=& h_{a,c}^{\cal R}(R)
\Phi^{{\cal R}}_{a,c}(\chi,w)\,, \nonumber\\
{\cal I}\{{\cal C}^{a,c}_3(\underline{\bm x})\}&=& h_{a,c}^{\cal I}(R)
\Phi^{{\cal I}}_{a,c}(\chi,w)\,,
\label{eq:shape&scaling}
\end{eqnarray}
where the function $h(R)$ is expected to have a scaling dependence in
the inertial range.

For the passive scalar, the scaling behavior of the reducible part,
$h^{\cal R}_c(R)$, is dominated by the dimensional scaling imposed by
the balance with the forcing, while the scaling of the irreducible
part, $h^{\cal R}_c(R)$, is given by the zero modes \cite{BGK98}. With
a finite-correlated pumping the forcing statistics may in principle
contribute to the irreducible part \cite{FGV01}; however in our case
these hypothetical contributions seem small, if not absent. For the
active scalar it is not possible to make any {\it a priori} argument
to predict the scaling behavior of the reducible and irreducible part.
As far as the geometrical dependence is concerned, $\Phi^{\cal R,I}_c$
and $\Phi^{\cal R,I}_a$, it is very difficult to separate the two
contributions. In our case, in agreement with the results obtained for
the Kraichanan model \cite{GLP97}, the reducible part turns out to be
the leading contribution, so that we only present the shape dependence
of the full correlation functions which basically coincides with the
reducible one.  Hereafter we will then use $\Phi_{a,c}$ dropping the
indices which distinguish the two contributions.

\begin{figure}[t!]
\vspace{-.2truecm}

\includegraphics[draft=false, scale=0.68, clip=true]{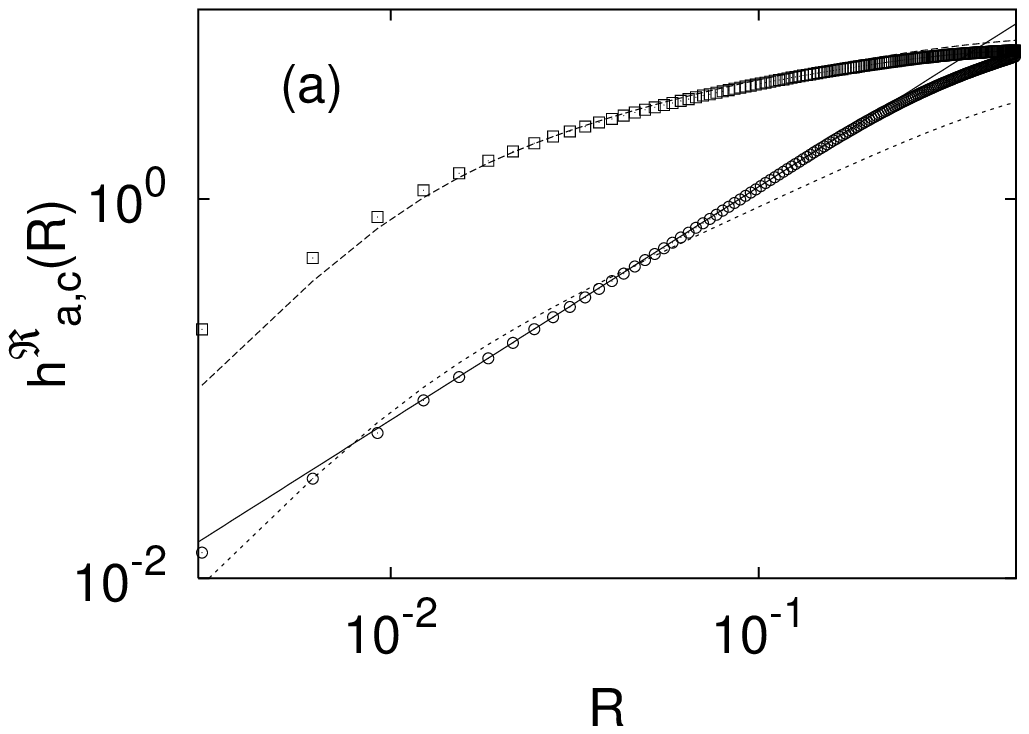} 
\includegraphics[draft=false, scale=0.68, clip=true]{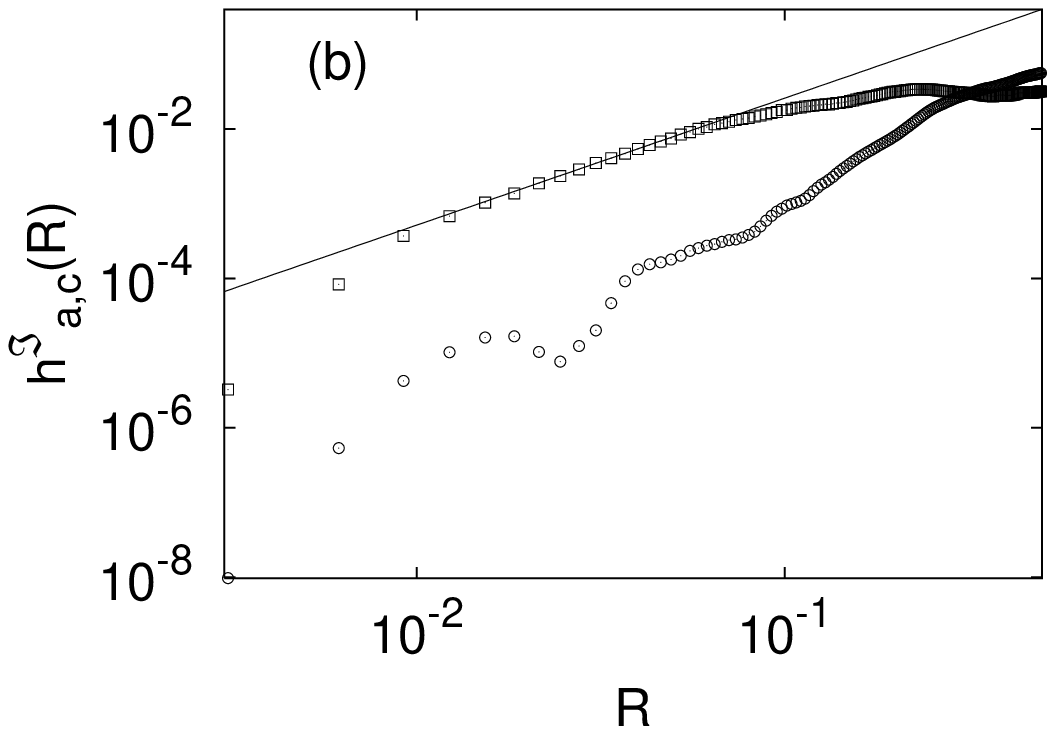} 
\caption{(a) $h^{\cal R}(R)$ vs $R$ for the active ($\circ$) and
passive ($\Box$) scalar. The dashed lines are the second order
structure functions, $S^{a,c}_2$, multiplied by a factor to allow for
a comparison.  The solid line gives the fitted slope, $R^{1.23\pm
0.05}$, for $h^{\cal R}_a(R)$.  The slope of $S^a_2$ is $\approx
r^{.84\pm 0.05}$ and is compatible with those obtained with the
forcing (G), see Fig.~\protect\ref{fig:phb3}).  The passive curves
have been shifted for visualization purposes. (b) $h^{\cal
I}_{a,c}(R)$ vs $R$ for the active ($\circ$) and passive ($\Box$)
scalar. For $c$ the scaling $R^{1.7\pm 0.2}$ (solid line) is
measured. The signal for the active scalar is very low and no scaling
behavior can be detected. 
The statistics has been computed averaging over
about $65$ frames separated by half eddy turnover time.}
\label{fig:phb6} 
\end{figure}

Let us start by studying the shape dependence of the
correlation function for the active, $\Phi_{a}(\chi,w)$, and passive
scalars, $\Phi_{c}(\chi,w)$. Exploiting the invariance under
arbitrary permutations of the three veritices of the triangle we can
reduce the configuration space to $\chi \in [-\pi/6:\pi/6]$, and $w\in
[0:1]$ going from degenerate (collinear points) to equilateral
triangles. The function $\Phi(\chi,w)$ is antiperiodic in $\chi$ with
period $\pi/3$ \cite{SS98,P98,CP01}. In Fig.~\ref{fig:phb5} the
functions $\Phi_c$ and $\Phi_c$ are shown. They have been measured for
a fixed size $R$ within the inertial range. 
The two functions display similar qualitative features: the
intensity grows going from equilateral to collinear triangles, and the
maximum is realized for almost degenerate triangles $(\chi,w)=(0,0)$.
$\Phi_c$ is invariant for $\chi \to -\chi$, which corresponds to
reflection with respect to an axis. This symmetry is a consequence of
the equation of motion and the fact that $f_c \to f_c$ under this
symmetry transformation. The active scalar displays a weak breaking of
this symmetry, as a consequence of the fact that it is a pseudoscalar
while $f_a$ is not.

Let us now turn our attention to the scaling behavior.  Since the
reducible scaling behavior is the leading one it is simply obtained by
fixing a certain shape for the triangle (we did for several choices of
$(\chi,w)$) and varying its size.  The results are shown in in
Fig.~\ref{fig:phb6}a. The observation is that $h^{\cal R}_a(R)$ and
$h^{\cal R}_c(R)$ are different.  Moreover, while $h^{\cal R}_c(R)$ is
practically parallel to $S^c_2(R)$, for the active scalar we observe
that $h^{\cal R}_a(R)$ is not scaling as $S^a_2(R)$. 

The measure of the subleading, irreducible part is more
involved, and we proceed as follows.  We fix a reference 
triangle shape $(\chi,w)$ and set the origin in the center of mass of the
triangle.
Now ${\bm x}_i$ indicates the position of the vertex $i$ of the 
triangle. We define $\hat{d}_1=\hat{d}_1(\lambda)$ ($\lambda \geq
1$) as the dilation operator which transforms the triangle $({\bm
x}_1,{\bm x}_2,{\bm x}_3)$ in $(\lambda {\bm x}_1,{\bm x}_2,{\bm
x}_3)$. Analogous definitions hold for the other vertices.
Obviously $\hat{d}_i(1)\equiv
\hat{I}$ is the identity.  We then consider the composite operator
$\hat{\cal D}(\lambda)=\hat{d}_1\hat{d}_2\hat{d}_3-
\hat{d}_1\hat{d}_2 -  \hat{d}_2\hat{d}_3 -\hat{d}_3\hat{d}_1
+\hat{d}_1+  \hat{d}_2 + \hat{d}_3 - \hat{I}$.  By
direct substitution it is easily seen that in $\hat{\cal D}(\lambda)
{\cal C}_3({\bm x}_1,{\bm x}_2, {\bm x}_3)$ all the reducible terms
 are cancelled, and only a linear superposition
of terms involving the irreducible parts survives
Therefore, the average of $\hat{\cal D}(\lambda)
c({\bm x}_1) c({\bm x}_2) c({\bm x}_3)$ for triangles of different
sizes $R$ but of fixed shape $(\chi,w)$ will give the scaling of the
irreducible part of the  three-point function.
(\ref{eq:shape&scaling}). A similar procedure has been done for $a$.  
We used as reference configuration a collinear triangle with two
degenerate vertices, i.e. $(\chi,w)=(0,0)$. This choice reduces the number of
cancellations.  Additionally, this configuration corresponds to
the region of stronger gradients in the function $\Phi$ (see
Fig.~\ref{fig:phb5}), yielding a higher signal to noise ratio. By
varying $\lambda$ we tested the robustness of the measured scaling.

In Fig.~\ref{fig:phb6} we present the results on the scale dependence
of $h^{\cal I}_c(R)$ and $h^{\cal I}_a(R)$.
Clearly $h^{\cal I}_a(R)$ has too low a signal to
identify any scaling behavior, while the passive scalar 
scales fairly well and we measured $h^{\cal I}_c(R)\sim R^{1.7}$,
confirming that the irreducible part is subleading.

A couple of final remarks are in order.
First, the fact that $h^{\cal R}_a(R)$ does not scale as $S^a_2(R)$ is the
signature of the correlations between $f_a$ and the
particle propagator. Indeed, for the passive scalar $h^{\cal
R}_c(R)\propto S^c_2(R)$ is a straightforward consequence 
of the independence of ${\bm v}$ and $f_c$.
Second, $h^{\cal R}_a(R)$ scales differently from $h^{\cal I}_c(R)$:
this rules out the possibility of establishing simple relationships
between active and passive scalar statistics
(see \cite{CCGP03} for a related discussion). 
\section{Conclusions and perspectives}
\label{sec:6}
In summary, we have investigated the statistics of active and passive
scalars transported by the same turbulent flow. We put the focus on
the issue of universality and scaling. In this respect the passive
scalar problem is essentially understood.  Conversely, the active case
is by far and away a challenging open problem.  The basic property
that make passive scalar turbulence substantially simpler is the
absence of statistical correlations between scalar forcing and carrier
flow. On the contrary, the hallmark of active scalars is the
functional dependence of velocity on the scalar field and thus on
active scalar pumping. Yet, when correlations are sufficiently weak,
the active scalar behaves similarly to the passive one: this is the
case of two-dimensional thermal convection and Ekman
turbulence. However, this appears to be a nongeneric situation, and
the equivalence between passive and active scalar is rooted in special
properties of those systems. Indeed, different systems as
two-dimensional magnetohydrodynamics or surface quasi-geostrophic
turbulence are characterized by a marked difference between passive
and active scalar statistics.  This poses the problem of universality
in active scalar turbulence: if the forcing is capable of influencing
the velocity dynamics, how can scaling exponents be universal with
respect to the details of the injection mechanism ? So far, a
satisfactory answer is missing.  As of today, numerical experiments
favor the hypothesis that universality is not lost in a number of
active scalars, but further investigation is needed to elucidate this
fundamental issue.
\ack

We are grateful to G.~Boffetta, S.~Musacchio, T.~Matsumoto for their
collaboration in part of the material here presented.  We thank
T.~Gilbert, A.~Noullez and I.~Procaccia for useful discussions.  This
work has been partially supported by the EU under the contracts
HPRN-CT-2000-00162 and HPRN-CT-2002-00300, and by Indo-French Center
for Promotion of Advanced Research (IFCPAR~2404-2).  AM and MC have
been partially supported by Cofin2003 (``Sistemi Complessi e Problemi
a Molti Corpi''). MC has been partially supported by the EU under the
contract HPRN-CT-2000-00162, and aknowledges the Max Planck Institute
for the Physics of Complex Systems (Dresden) for computational
resources.  Numerical simulations have been performed at IDRIS
(project 021226) and at CINECA (INFM parallel computing initiative).


\end{document}